\documentclass[12pt,preprint]{aastex61}
\usepackage{natbib}
\usepackage{hyperref} 
\usepackage{amsmath}
\newcommand{\um}{$\mu$m}

\newcommand{\Msun}{M$_{\odot}$}

\newcommand{\Lsun}{L$_{\odot}$}

\newcommand{\kms}{km~s$^{-1}$}

\newcommand{\cms}{${\rm cm}^{-2}$}
\newcommand{\cmc}{${\rm cm}^{-3}$}
\newcommand{\pcc}{${\rm cm}^{-3}$}
\newcommand{\hi}{\mbox{$\mathrm{H\,{\scriptstyle {I}}}$}}
\newcommand{\hii}{\mbox{$\mathrm{H\,{\scriptstyle {II}}}$}}

\newcommand{\vlsr}{V$_{LSR}$}

\newcommand{\nthpnt}{N$_{2}$H$^{+}$}

\newcommand{\nht}{NH$_{3}$}
\newcommand{\hcopnt}{HCO$^{+}$}
\newcommand{\hcop}{HCO$^{+}$\,(1--0)}

\newcommand{\tastar}{T$^{*}_{A}$}

\newcommand{\cii}{[\mbox{$\mathrm{C\,{\scriptstyle {II}}}$}]}
\newcommand{\tcii}{[\mbox{$^{13}\mathrm{C\,{\scriptstyle {II}}}$}]}
\newcommand{\oi}{[\mbox{$\mathrm{O\,{\scriptstyle {I}}}$}]}

\def\addtodot#1.#2\relax{#1\rlap{.}^{\dotadd}#2}
\newcommand{\dotdeg}[1]{\def\dotadd{\circ}\addtodot#1\relax}

\shorttitle{ \cii\, and \oi\, Absorption and Self-absorption in the Nessie IRDC}
\shortauthors{Jackson et al.}

\begin{document}

\title{Absorption and Self-Absorption of \cii\, and \oi\, Far Infrared Lines Towards a Bright Bubble in the Nessie Infrared Dark Cloud}

\author{James M. Jackson}
\affiliation{Green Bank Observatory,155 Observatory Road, Green Bank, WV 24944, USA}
\affiliation{USRA SOFIA Science Center, NASA Ames Research Center, Moffett Field, CA 94045, USA}
\affiliation{School of Mathematical and Physical Sciences, University of Newcastle, University Drive, Callaghan NSW 2308, Australia}

\author{J. Scott Whitaker}
\affiliation{Physics Department, Boston University, 590 Commonwealth Ave., Boston, MA 02215, USA}

\author{Edward Chambers}
\affiliation{USRA SOFIA Science Center, NASA Ames Research Center, Moffett Field, CA 94045, USA}

\author{Robert Simon}
\affiliation{I. Physikalisches Institut, Universität zu Köln, Zülpicher Str. 77, 50937 Köln, Germany}

\author{Cristian Guevara}
\affiliation{I. Physikalisches Institut, Universität zu Köln, Zülpicher Str. 77, 50937 Köln, Germany}

\author{David Allingham}
\affiliation{School of Mathematical and Physical Sciences, University of Newcastle, University Drive, Callaghan NSW 2308, Australia}

\author{Philippa Patterson}
\affiliation{School of Mathematical and Physical Sciences, University of Newcastle, University Drive, Callaghan NSW 2308, Australia}

\author{Nicholas Killerby-Smith}
\affiliation{School of Mathematical and Physical Sciences, University of Newcastle, University Drive, Callaghan NSW 2308, Australia}

\author{Jacob Askew}
\affiliation{School of Mathematical and Physical Sciences, University of Newcastle, University Drive, Callaghan NSW 2308, Australia}

\author{Thomas Vandenberg}
\affiliation{School of Mathematical and Physical Sciences, University of Newcastle, University Drive, Callaghan NSW 2308, Australia}

\author{Howard A. Smith}
\affiliation{Center for Astrophysics, Harvard \& Smithsonian, 60 Garden St., Cambridge, MA 02138, USA}

\author{Patricio Sanhueza}
\affiliation{National Astronomical Observatory of Japan, National Institute of Natural Sciences, 2-21-1 Osawa, Mitaka, Tokyo 181-8588, Japan}
\affiliation{Astronomical Science Program, The Graduate University for Advanced Studies, SOKENDAI, 2-21-1 Osawa, Mitaka, Tokyo 181-8588, Japan}

\author{Ian W. Stephens} 
\affiliation{Center for Astrophysics, Harvard \& Smithsonian, 60 Garden St., Cambridge, MA 02138, USA}
\affiliation{Department of Earth, Environment, and Physics, Worcester State University, 486 Chandler Street, Worcester MA  01602, USA}

\author{Lars Bonne}
\affiliation{USRA SOFIA Science Center, NASA Ames Research Center, Moffett Field, CA 94045, USA}

\author{Fiorella L. Polles}
\affiliation{USRA SOFIA Science Center, NASA Ames Research Center, Moffett Field, CA 94045, USA}

\author {Anika Schmiedeke}
\affiliation{Green Bank Observatory,155 Observatory Road, Green Bank, WV 24944, USA}

\author {Netty Honigh}
\affiliation{I. Physikalisches Institut, Universität zu Köln, Zülpicher Str. 77, 50937 Köln, Germany}

\author{Matthias Justen}
\affiliation{I. Physikalisches Institut, Universität zu Köln, Zülpicher Str. 77, 50937 Köln, Germany}

\begin{abstract}
Using the upGREAT instrument on SOFIA, we have imaged \cii\, 157.74 and \oi\, 63.18 micron line emission from a bright photodissociation region (PDR) associated with an ionized ``bubble'' located in the Nessie Nebula, a filamentary infrared dark cloud. A comparison with ATCA data reveals a classic PDR structure, with a uniform progression from ionized gas, to photodissociated gas, and on to molecular gas from the bubble's interior to its exterior. \oi\, line emission from the bubble's PDR reveals self-absorption features. Toward a FIR-bright protostar, both \oi\, and \cii\, show an absorption feature at a velocity of $-18$ \kms, the same velocity as an unrelated foreground molecular cloud. Since the gas density in typical molecular clouds is well below the \oi\, and \cii\, critical densities, the excitation temperatures for both lines are low ($\sim 20$ K). The Meudon models demonstrate that the surface of a molecular cloud, externally illuminated by a standard $G_0 = 1$ interstellar radiation field, can produce absorption features in both transitions. Thus, the commonly observed \oi\, and \cii\, self-absorption and absorption features plausibly arise from the subthermally excited, externally illuminated photodissociated envelopes of molecular clouds. The luminous young stellar object AGAL337.916-00.477, located precisely where the expanding bubble strikes the Nessie filament, is associated with two shock tracers: \nht\, (3,3) maser emission and SiO $2-1$ emission, indicating interaction between the bubble and the filament. The interaction of the expanding bubble with its parental dense filament has triggered star formation.
 \end{abstract}

\section{Introduction}

The \cii\, 157.74 \um\, $^2P_{3/2}$--$^2P_{1/2}$ line and the \oi\, 63.18 \um\, $^3P_1$--$ ^3P_2$ line are important probes of the physical state of gas in photodissociation regions (PDRs; also known as ``photon dominated regions''), the interface between photoionized gas and molecular gas surrounding recently formed OB stars. Since these two lines are typically the dominant cooling lines of PDRs, they are very luminous, and their fluxes are key inputs to PDR models (e.g., \citealt{TielensHollenbach1985, Kaufman2006, Pound2008}).  \cii\, emission is also commonly used to indicate the global star-formation rate of galaxies (e.g., \citealt{Stacey2010}).  Current diagnostic PDR models, however, rarely take into account the possibility of foreground gas that might absorb the line emission from the PDR or the continuum emission from an embedded young stellar object.  

Many studies, however, show that for both the \cii\, and \oi\, lines foreground gas often produces absorption features toward embedded far infrared (FIR) continuum sources, or self-absorption dips against background PDR line emission (e.g., \cii: \citealt{Vastel2002, Gerin2015, Graf2015, Guevara2020, Kirsanova2020, Jacob2022}, \oi: \citealt{Poglitsch1996, Kraemer1998, Vastel2000, Lis2001, Vastel2002, Gerin2015, Wiesemeyer2016, Okada2019, Goldsmith2021, Jacob2022, Goldsmith2023}).  For \cii, optical depths sufficient for absorption ($\tau \sim 1$) have been established by comparing the \cii\, flux with the optically thin \tcii\, hyperfine line fluxes (e.g., \citealt{Okada2019, Guevara2020}) or by comparing ``on'' and ``off'' beam spectra \citep{Gerin2015}. Evidence for sufficiently large optical depths to produce absorption or self-absorption in the \oi\, 63 \um\, line comes from both direct observation of self-absorption or absorption features (e.g., \citealt{Poglitsch1996, Gerin2015, Wiesemeyer2016, Goldsmith2021, Goldsmith2023}), and also the lower-than-expected observed intensity ratio between the 
63 \um\, $^3P_1$--$ ^3P_2$ and the 145 \um\, $^3P_0$--$ ^3P_1$  \oi\, lines (e.g., \citealt{Kraemer1998, Liseau2006, Abel2007, Ossenkopf2015}).  \cite{Goldsmith2021,Goldsmith2023} explicitly model \oi\, self-absorption and conclude that self-absorption reduces the \oi\ integrated instensity from the PDR in W3 by factors of 2 to 4.  \rm In this paper, we examine \cii\, and \oi\, line observations, as well as cm and mm wave radio observations, toward a bubble-like \hii\, region in the ``Nessie Nebula,'' and detect both absorption and self-absorption features in the \oi\, line, and absorption features in the \cii\, line.

The ``Nessie Nebula'' is a filamentary Infrared Dark Cloud (IRDC) first identified by \cite{Jackson2010} from the {\it Spitzer}/GLIMPSE mid-IR images of the Galactic plane \citep{Benjamin2003,Benjamin2005}. Molecular line observations \citep{Jackson2010} established that the Nessie Nebula is a coherent, filamentary structure spanning at least 1.5 degrees in Galactic longitude, at a kinematic distance of approximately 3.1 kpc \citep[see also ][]{Goodman2014}. This IRDC is both unusually long, with a length of at least 80 pc, and also unusually elongated, with an aspect ratio of at least 150:1 \citep{Jackson2010} and probably  larger \citep{Goodman2014}.  The filament has a linear mass density (mass per unit length) of 627 \Msun\, pc$^{-1}$ \citep{Mattern2018} and contains 16 dense molecular clumps \citep{Csengeri2014} identified from the ATLASGAL survey \citep{Schuller2009}.  With masses ranging from $\sim 10^2$ to $10^3$ \Msun\, \citep{Mattern2018}, these clumps are the current or future sites of high-mass star-formation.  Based on the evolutionary classification scheme described in \cite{Jackson2013}, these clumps span a wide range of evolutionary stages, from the earliest pre-stellar or ``quiescent'' stage, through the intermediate ``protostellar'' stage, and on to the latest``\hii\, region'' stage. 

The most active star-forming region in the Nessie Nebula is associated with a shell-like \hii\, region ``bubble,'' hereafter referred to as the ``Nessie Bubble,'' centered near Galactic coordinates $(l,b) = (\dotdeg{337.97},\dotdeg{-0.47})$.  This tear-drop shaped shell is bright in dust continuum emission as well as molecular line emission (e.g., \citealt{Jackson2010}). It was designated S36 in the GLIMPSE study of Galactic Plane \hii\ region bubbles \citep{Churchwell2006}.  The Nessie Bubble is most luminous on its western edge.       This region had been previously noted as the site of a candidate stellar cluster, DBS2003-157 \citep{Dutra2003}.  Indeed,  \cite{Messineo2018} identified seven candidate OB stars in close projected proximity to the western edge of the Nessie Bubble. This region is associated with a compact 100-150 GHz radio continuum source, GRS G337.92-00.48 \citep{Culverhouse2011}, an IRAS far-infrared (FIR) continuum source, 16374-4701 \citep{Culverhouse2011}, and an ATLASGAL \citep{Schuller2009} 870 \um\, continuum source, AGAL337.916-00.477  \citep{Contreras2013}.  This luminous young stellar object (YSO; $L= 50,000$ \Lsun) hosts a powerful bipolar outflow detected in CO $1-0$ and $3-2$ \citep{Torii2017}.   \cite{Motte2022} estimate a total gas mass of $2.5 \times 10^3$ \Msun\, for a $\sim 1 \times 1$ pc region centered on the continuum source, and resolve the region into numerous cores, the most massive of which contains 160 \Msun.  

Because the Nessie Bubble is ionized internally by the ultraviolet radiation from OB stars and associated with molecular gas, it is likely to contain a PDR, the transition zone from completely ionized to completely molecular gas (e.g., \citealt{TielensHollenbach1985}).   This paper describes new SOFIA far-infrared observations, using the upGREAT instrument, Australia Telescope Compact Array 1.3 cm observations, and the 22 m Mopra telescope 3 mm observations of the western portion of the Nessie Bubble.  The data reveal the PDR structure of the Nessie Bubble and show complex \cii\, and \oi\, line profiles, with absorption and self-absorption features.  This paper explores the hypothesis that a standard interstellar radiation field that externally illuminates molecular clouds can generate both the low excitation temperature as well as sufficient optical depth necessary to produce \cii\, and \oi\, absorption and self-absorption features by creating a low-density, subthermally excited PDR on the cloud's periphery.  Detailed modeling described in this paper confirms the viability of this hypothesis.

Two distinct shock tracers, \nht\, (3,3) maser emission and SiO $2-1$ thermal emission, indicate that the most luminous YSO in the region occurs at the location where the Nessie Bubble is interacting with the Nessie IRDC filament.  This location strongly suggests that the bubble-filament interaction triggered the YSO's formation. Finally, this paper explores a speculative evolutionary scenario where successive interactions between the expanding Nessie Bubble and the Nessie IRDC filament have produced a wave of propagating star-formation that proceeds along the filament and produces many of the observed structures.  

\section{Observations}

\subsection{SOFIA Observations}  In 2017, we used SOFIA with the upGREAT array receiver to make simultaneous pointed observations of the \oi\, 63 \um\, and \cii\, 158 \um\,  lines centered on the ATLASGAL continuum peak AGAL337.916-00.477.    SOFIA, the Stratospheric Observatory For Infrared Astronomy, is described in \cite{Young2012} and  \cite{Temi2018}.  The upGREAT instrument \citep{Risacher2018} employs heterodyne receivers in hexagonal arrays.  Two seven-element arrays cover the low frequency range with dual polarization, allowing observations of the 158 \um\, \cii\, line; a single seven-element array covers the high frequency range for observations of 63 \um\, \oi.
In 2018 we used SOFIA/upGREAT to map a $\sim 0.10 \times 0.05$ deg region of the western edge of the Nessie Bubble using an on-the fly raster mapping technique.  Due to the shorter intergration times on each position in the map, these mapping data have lower S/N than the 2017 pointed observations.  The FWHM beam size is 14.1 arcsec for the \cii\, line and 6.3 arcsec for the \oi\, line.

\subsection{ATCA Observations}
The western portion of the Nessie Bubble was observed with the Australia Telescope Compact Array (ATCA), near Narrabri, NSW, Australia using the 15 mm receiver and the Compact Array BroadBand (CABB) backend \citep{Wilson2011}.  These observations are part of a larger survey project, the Complete ATCA Census of High-Mass Clumps \citep{Allingham2024}, which has imaged sixty dense molecular clumps using the NH$_3$ (1,1) through (6,6) inversion lines from 23.6945 GHz through 25.05602 GHz, the H$_2$O maser line at 22.23508 GHz, and several other 1.3 cm lines, mostly hydrogen recombination lines and CH$_3$OH lines.  In addition, 22.18 and 24.06 GHz continuum images were generated.  For the Nessie Bubble observations, the field center was $(l,b) = (\dotdeg{337.916},\dotdeg{-0.477})$, the position of AGAL337.916-00.477.  Observations were taken between 20 July 2017 and 21 June 2020 in 5 different array configurations:  H75, H168, H214, 750C, and 1500C.  To improve $u,v$ coverage and sensitivity, the Nessie Bubble was observed in at least two separate observing sessions for each array configuration.  The flux calibrator was 1934-638, the bandpass calibrator 1235-055, and the phase calibrator 1613-586.  The $u,v$ data were combined, edited, imaged, and CLEANed using standard techniques.  This paper presents images of NH$_3$ (1,1) (rest frequency 23.6944955 GHz) and (3,3) (rest frequency 23.8701292 GHz) with a synthesized beam size of $2.7'' \times 3.6''$, an rms sensitivity of 0.45 mJy/beam, and a spectral resolution of 0.4 \kms.  The conversion between fluxes in mJy beam$^{-1}$ and brightness temperature in K is 1 mJy = 0.22 K for both transitions.  Details on the observations and reduction are presented in \cite{Allingham2024}.

\subsection{Mopra Observations}
We imaged the entire Nessie IRDC, including the Nessie Bubble, with the 22 m Mopra telescope near Coonabarabran, NSW, Australia in several molecular lines in the 86-93 GHz frequency range using the MOPS backend.  The observations were conducted from 7 May 2018 to 13 September 2018.  The observing set up  was identical to that used for the MALT90 Survey \citep{Jackson2013}.  Here we present data for the SiO $2-1$ line, a standard shock tracer, with a rest frequency of 86.846960 GHz.  The FWHM beam size is 75$''$; the typical rms sensitivity is 0.035 Jy; and the spectral resolution is 0.11 \kms. 

\section{Results}

\begin{figure}
\begin{center}
 \includegraphics[scale=1.0, angle=0]{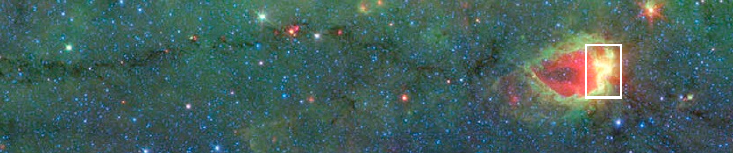} 
\includegraphics[scale=0.90, angle=0]{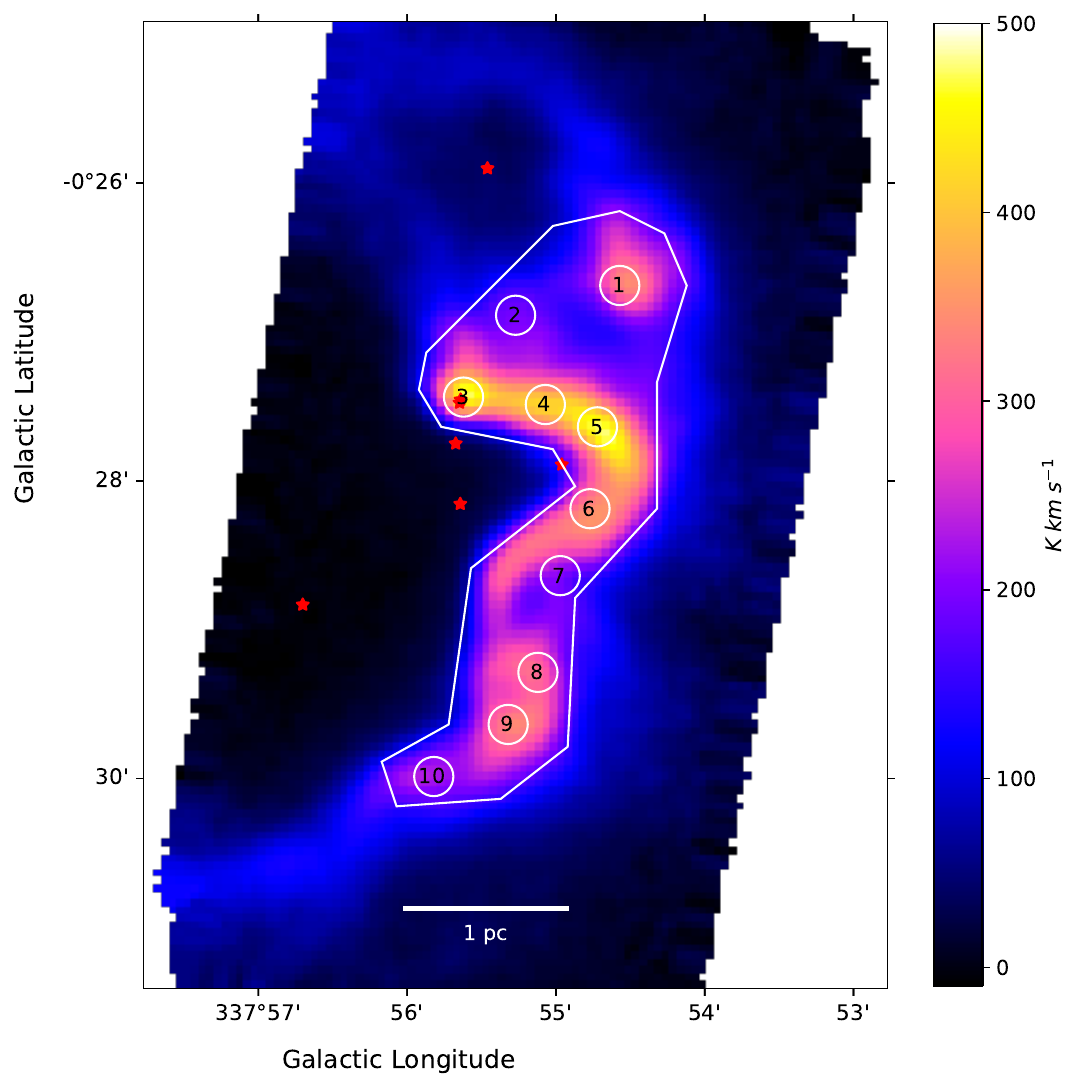} 
\caption{ Top: A Spitzer IRAC/MIPS image of a portion of the Nessie IRDC  \citep{Jackson2010}  with 24 \um\, in red, 8 \um\, in green, and 3.6 \um\, in blue.  The image measures approximately 1.1 degrees in Galactic longitude by 0.25 degrees in Galactic latitude.  Our current study addresses the Nessie-A region within the white rectangle.  Bottom: A map of the \cii\, integrated intensity over the velocity range  of -39.8 $\pm$4.4\kms\, (-44.2 to -35.4 \kms). The red star symbols indicate the positions of OB stars as listed in Table 2 of \citealt{Messineo2018}.  The white polygon  marks out the Nessie-A region for further analysis, and the numbered circles designate spots where spectra will be reported.  Spot 7 coincides with the peak of 870 \um\, emission reported by \citealt{Schuller2009} and designated AGAL337.916-00.477 \citep{Contreras2013}.}
\label{fig:figcii}
\end{center}
\end{figure}

\begin{figure}
\begin{center}
\includegraphics[scale=1.0, angle=0]{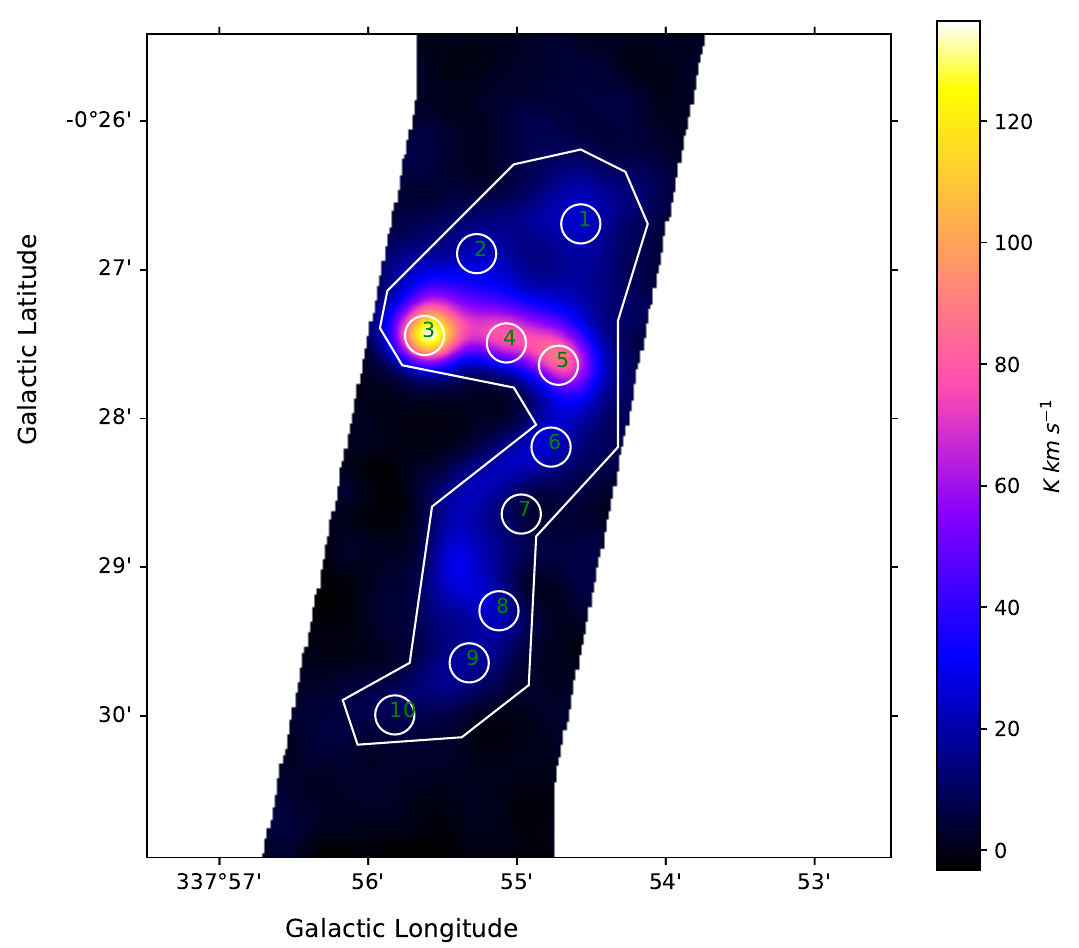} 
\caption{ A map of  the \oi\, 63 \um\, intensity integrated over the velocity range  of -39.8 $\pm$4.4\kms .  The \oi\ data have been smoothed and regridded to the same resolution as  the \cii\, map in Figure \ref{fig:figcii}.  The numbered circles are the same as in Fig. \ref{fig:figcii} and indicate spots where spectra will be reported.}
\label{fig:figoi}
\end{center}
\end{figure}

\begin{figure}
\begin{center}
\includegraphics[scale=0.85, angle=0]{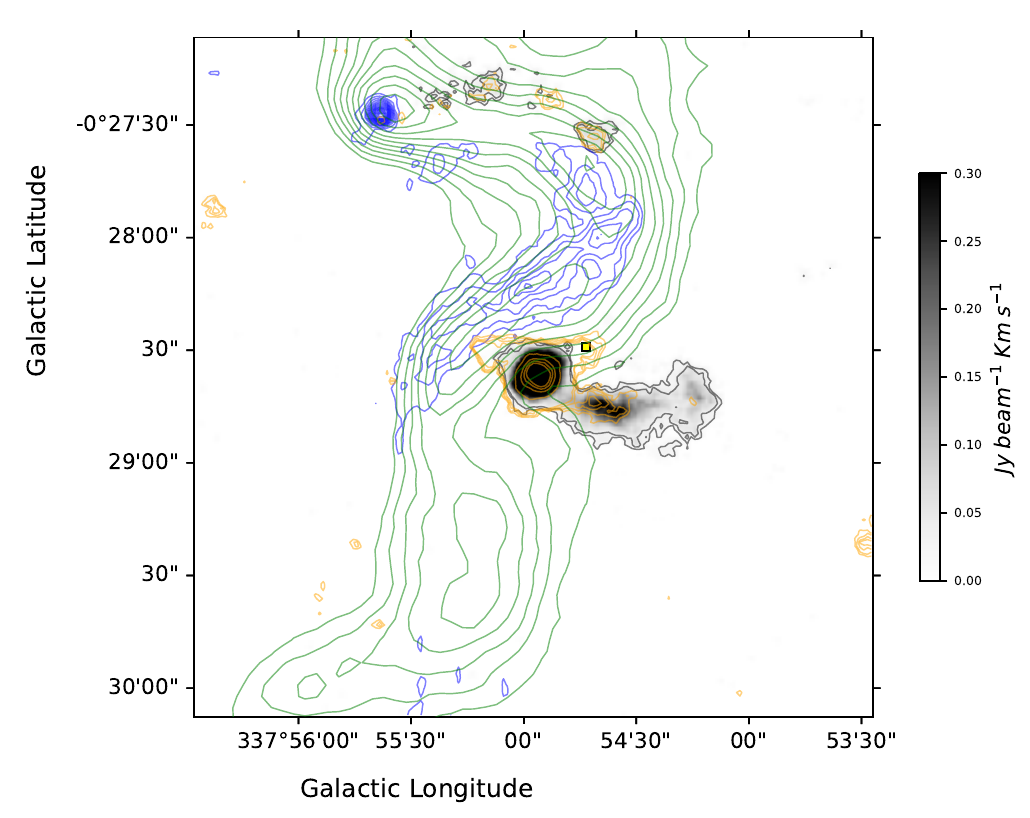} 
\caption{Maps of the western portion of the Nessie Bubble using an ionized gas tracer (blue contours: 24 GHz radio continuum), a PDR tracer (\cii: green contours), a colder molecular gas tracer (\nht\, (1,1): gray scale and gray contours), and a warmer molecular gas tracer (\nht\, (3,3): gold contours).   The 24 GHz continuum is shown in blue with contours at 2 to 50 mJy beam$^{-1}$ in steps of 2 mJy beam$^{-1}$. \cii\, integrated emission is indicated by the green contours from 200 to 700 K \kms\, in steps of 50 K \kms.The \nht\, (3,3) contour levels are 3.,  5., 7.5, 10., 50., 100, 150, 200, 500, 1000, 1500, 2000, 2500, and 3000 mJy beam$^{-1}$ \kms.  The yellow/black square shows the location of the \nht\, (3,3) maser.}
\label{fig:figmapall}
\end{center}
\end{figure}

Figure \ref{fig:figcii} shows in the upper panel a Spitzer IRAC/MIPS 3.6, 8.0, and 25 \um\, image of the Nessie IRDC and in the lower panel the new SOFIA integrated intensity map of the \cii\, emission from the western edge of the Nessie Bubble taken with the upGREAT instrument.  The white polygon in Figure \ref{fig:figcii} marks the region over which average spectra will be calculated; the numbered circles denote specific positions in the Bubble which are the subject of more detailed analysis below.   The bright ATLASGAL continuum source AGAL337.916-00.477 \citep{Contreras2013} coincides with Spot 7 in Figure \ref{fig:figcii}, and candidate OB stars are marked by red star symbols \citep{Messineo2018}.
Figure \ref{fig:figoi}  presents the SOFIA map of  the \oi\,  intensity integrated over the velocity range  of $-39.8\pm4.4$ \kms .  The \oi\ data have been smoothed and regridded to the same resolution as  the \cii\, map in Figure \ref{fig:figcii}.  
Here and throughout, we report the \cii\, and \oi\, intensities on the Rayleigh-Jeans brightness temperature scale $T_{B,RJ} = {{I_\nu \lambda^2}\over{2k}}$, where $I_\nu$ is the intensity, $\lambda$ the wavelength, and $k$ the Boltzmann constant. 
The conversion between intensity and brightness temperature is 1 K $=$ 588 Jy for \cii\, and 1 K $=$ 731 Jy for \oi.
Figure \ref{fig:figmapall} presents a map of the 24 GHz radio continuum flux density (blue), as observed with the ATCA interferometer, the \cii\, emission (green) from SOFIA, as well as \nht\, (1,1) (gray) and (3,3) (gold) maps from ATCA.   Figure \ref{fig:figposprofiles} shows spectra for \cii\, and \oi\, at ten points along the major \cii\, emission pattern with positions indicated in Figures \ref{fig:figcii} and \ref{fig:figoi}.  The green lines are Gaussian fits to the [C II] spectra between the shorter vertical dashed lines.

\begin{figure}
\begin{center}
\includegraphics[scale=0.7, angle=0]{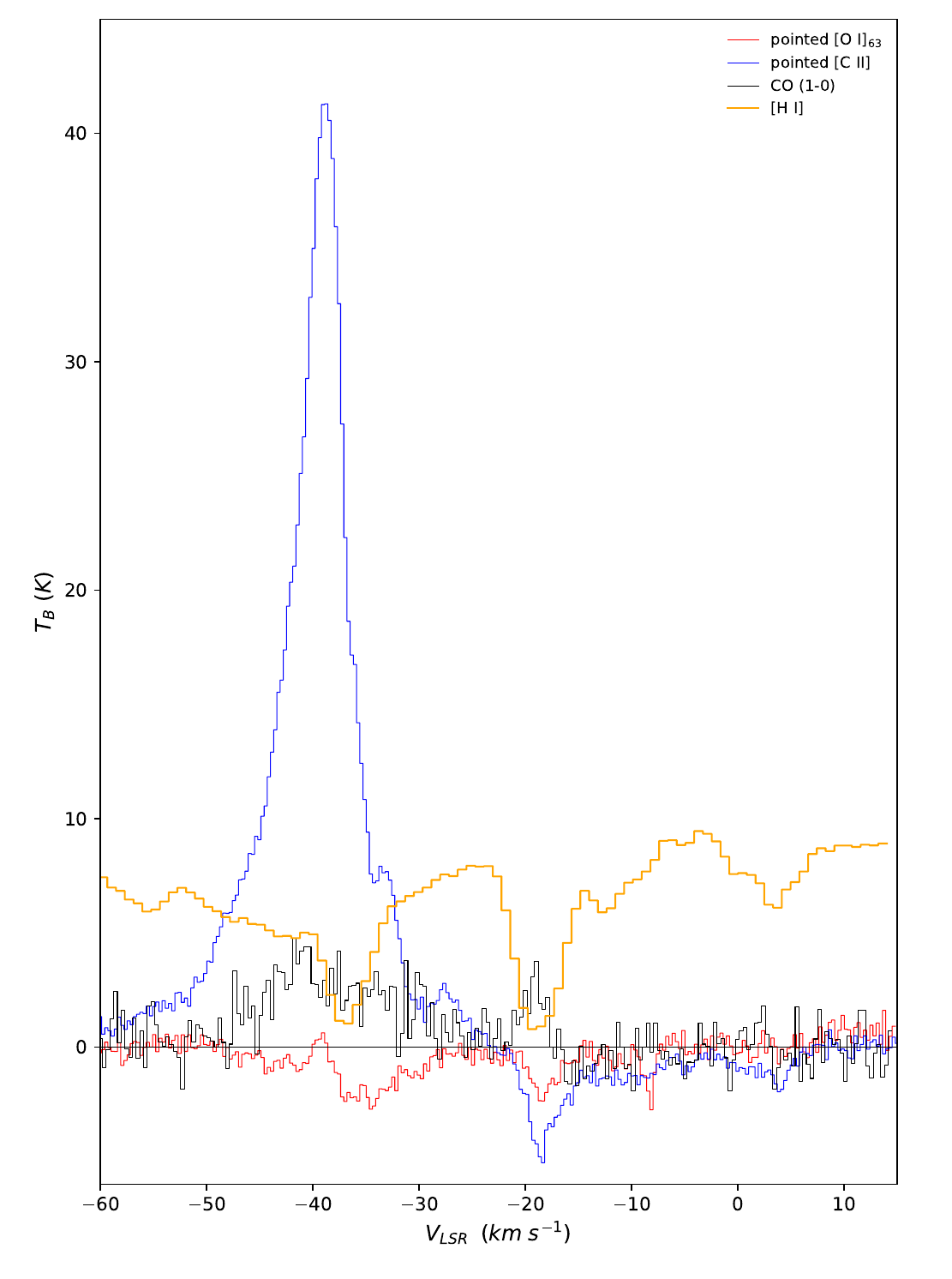} 
\caption{Spectra towards the luminous protostar AGAL337.916-00.477.  Blue: \cii, red: \oi, black: CO $1-0$ \citep{Barnes2015}, and gold: \hi\, \citep{McClureGriffiths2005}.  The \cii\, and \oi\, spectrum are from the deep integration single-pointing data from SOFIA Cycle 5.  For the \hi\, spectrum, the continuum level has been retained.  For all other spectra the continuum baseline has been removed.  At $V_{LSR} = -35$ \kms, both \oi\, and \hi\, show absorption features against the continuum.  At $V_{LSR} = -18$ \kms, \cii, \oi, and \hi\, show an absorption feature, while the CO shows an emission feature.  At $V_{LSR} = +3$ \kms, \hi\, shows an absorption feature, corresponding to a marginal absorption feature in \cii\, and emission feature in CO. }
\label{fig:figcontprofiles}
\end{center}
\end{figure}

\begin{figure}
\begin{center}
\includegraphics[scale=0.7, angle=0]{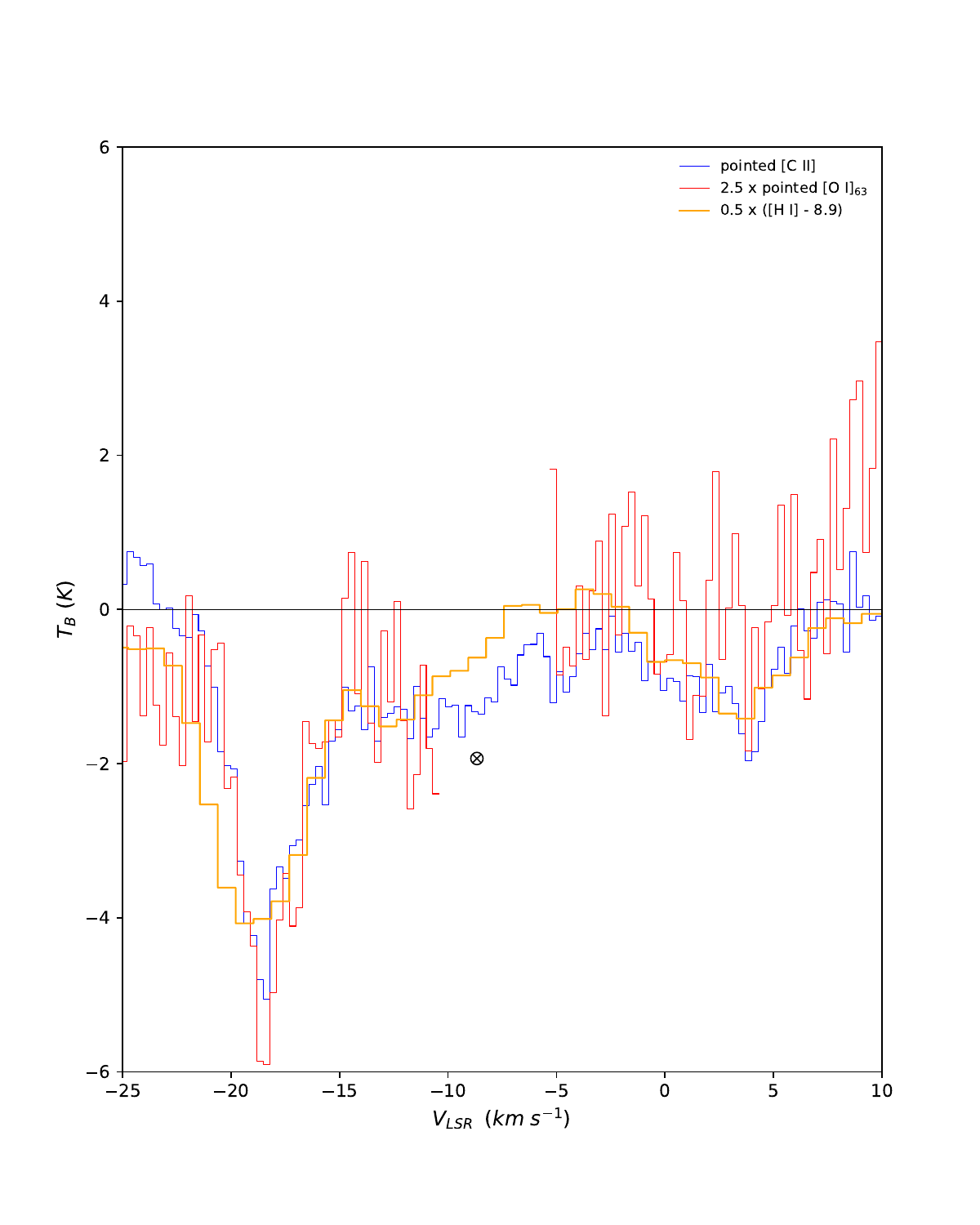}
\caption{A  close-up view of the continuum-subtracted \cii\, (blue), \oi\, (red), and \hi\, (gold) spectra over the velocity range $V_{LSR} = -25$ to 10 \kms\,  towards the luminous protostar AGAL337.916-00.477.  The line profiles are essentially identical within the noise. The \hi\, and \oi\, spectra have been scales by factors of 0.5 and 2.5, respectively.}
\label{fig:specshapes}
\end{center}
\end{figure}

Figure \ref{fig:figcontprofiles} shows the \cii\, (blue) and \oi\, (red) spectra toward AGAL337.916-00.477 from the deeper, single-pointing 2017 SOFIA/upGREAT observations, as well as the CO $1-0$ spectrum (black) from the Three-mm Ultimate Mopra Milky Way Survey (ThrUMMS, \citealt{Barnes2015}) and the 21 cm \hi\, (gold) spectrum from the Southern Galactic Plane Survey (SGPS) data \citep{McClureGriffiths2005} .  In addition to the superposed narrow emission and broad absorption features  at $-40$ \kms,  the \oi\, spectrum also exhibits a clear absorption dip at $-18$ \kms.  The \cii\, spectrum also shows a deep absorption feature at $-18$ \kms. We interpret these features as arising from absorption in an intervening, foreground cloud.  This interpretation is supported by the presence at the same velocity of an emission feature in CO and an absorption feature in \hi.  Gaussian fits to this \cii\, absorption feature and the corresponding CO emission feature yield LSR velocities of -18.2 and -19.1 \kms, respectively.    There is marginal evidence for an additional absorption feature in \cii\ at $\sim +3$ \kms. Unfortunately, since the \oi\, spectrum is contaminated by telluric \oi\, at this velocity, we are unable to confirm the presence of \oi\, absorption at +3 \kms. Both \cii\, absorption features at $-18$ and $+3$ \kms\, are coincident with \hi\, 21 cm absorption features  and CO emission features, evident in both in the CO $1-0$ ThrUMMS data and also  the CO $1-0$ and $3-2$ spectra presented by \cite{Torii2017}.  Figure \ref{fig:specshapes} presents a detailed comparison of the \hi, \cii, and \oi\, absorption profiles from $V_{LSR} = -25$ to 10 \kms.  The line shapes are remarkably similar, and indicate that each line is tracing the same gas.

\begin{figure}
\begin{center}
\includegraphics[scale=0.85, angle=0]{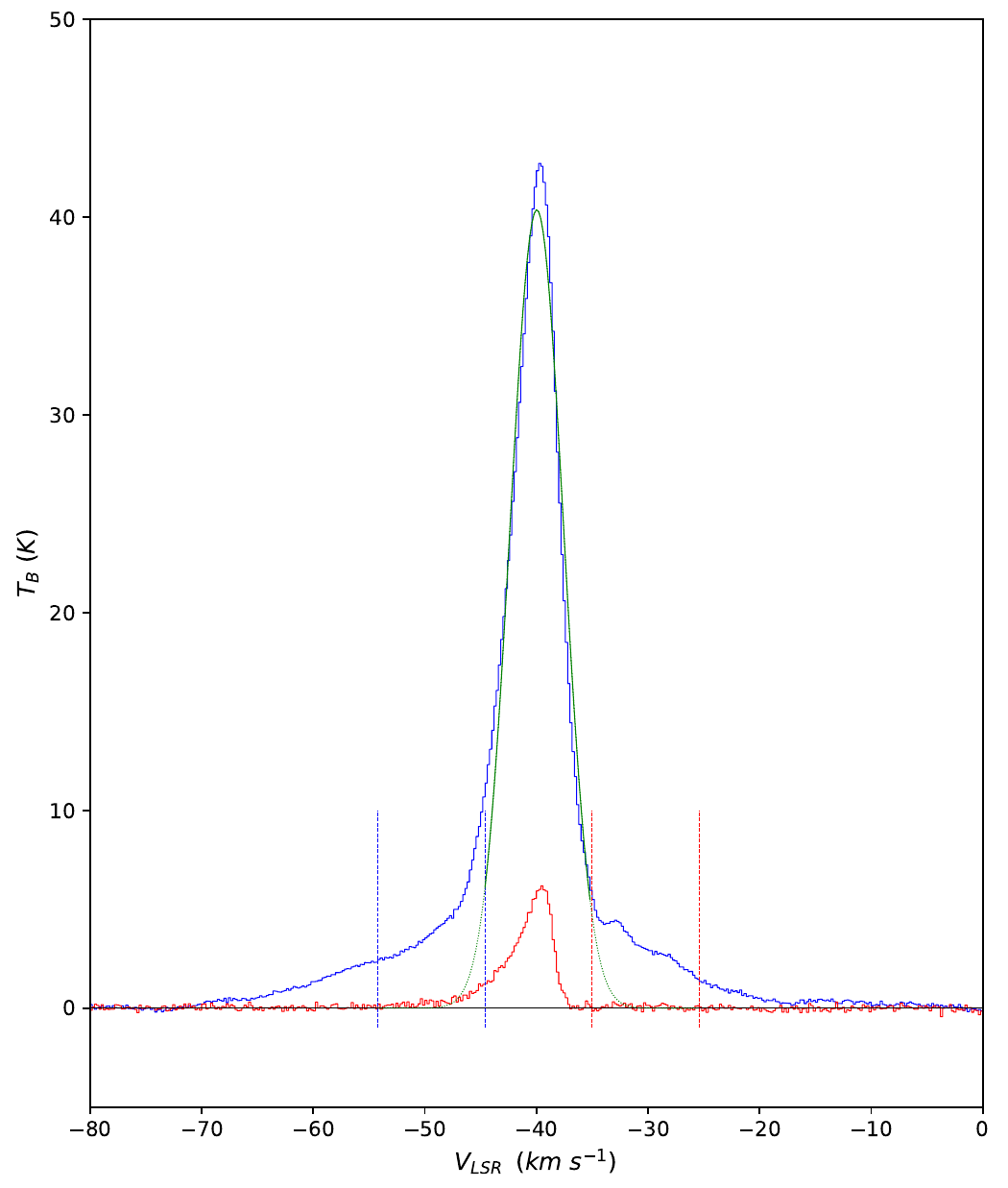} 
\caption{Average spectra of \cii\, (blue) and \oi\, (red) over the pixels enclosed by
the polygon in Figure \ref{fig:figcii}. The vertical scale shows $T_{MB}$, the Rayleigh-Jeans main-beam brightness temperature in Kelvin. The solid
green curve is a Gaussian fit to the central portion of the [C II] spectrum, extended as the dashed green line. This Gaussian has a velocity dispersion $\sigma_V$ of 2.4 km/s.  Note the asymmetric shape of the \oi\, profile that likely indicates self-absorption on the red-shifted side of the background emission line.  The vertical dashed lines indicate the Blue, Central, and Red emission
ranges used in Figure \ref{fig:ciiredblue}.  Each range has a width of four times the dispersion of the central Gaussian fit.
}
\label{fig:figavgprofiles}
\end{center}
\end{figure}

\begin{figure}
\begin{center}
\includegraphics[scale=0.85, angle=0]{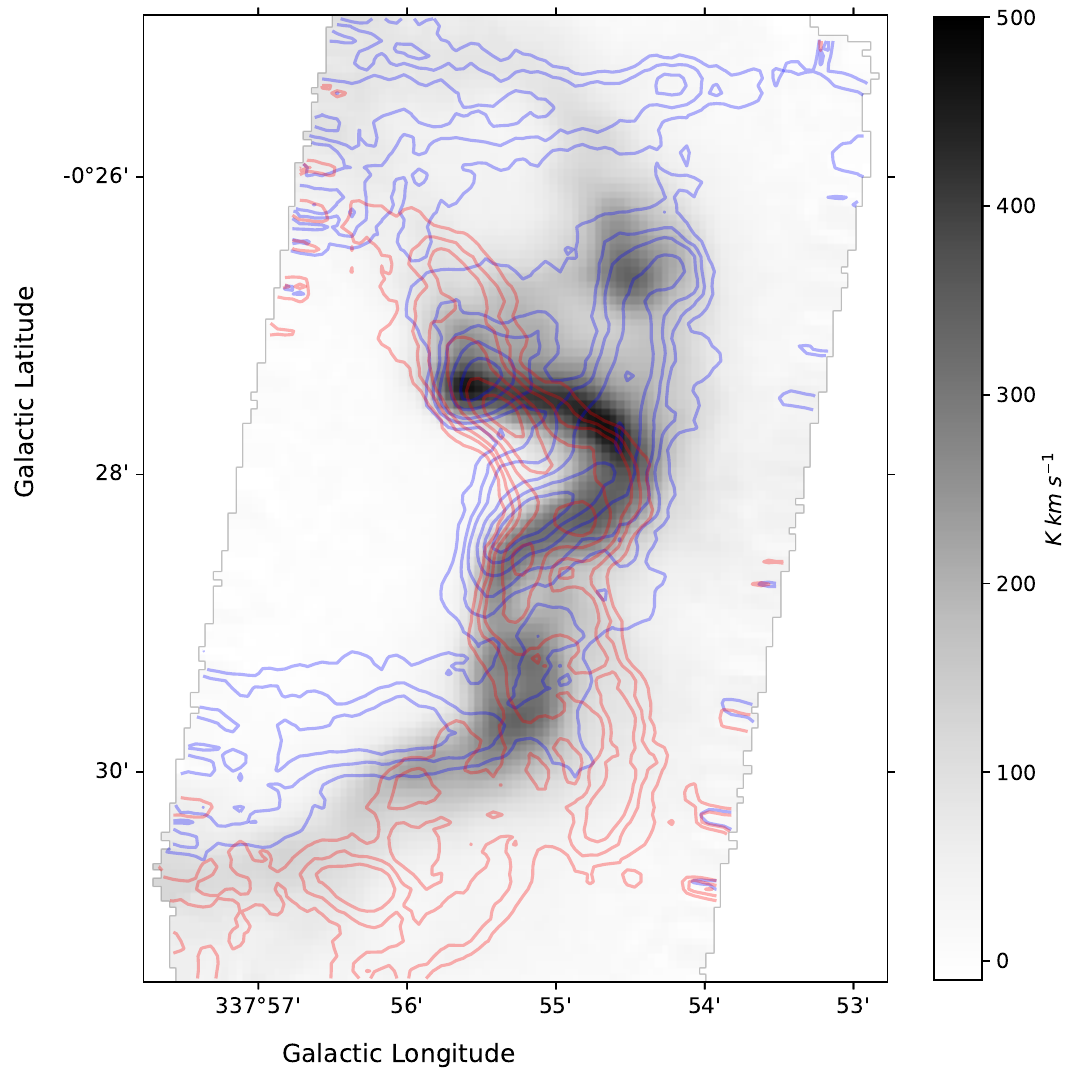} 
\caption{ The grayscale image shows the \cii\, intensity integrated over the Central velocity range in Figure \ref{fig:figavgprofiles}.  The blue(red) contours show the \cii\, integrated intensity in the Blue(Red) velocity ranges indicated in Figure \ref{fig:figavgprofiles}.  The contours are at 20, 30, 40, 60, 80, and 100 K \kms.}
\label{fig:ciiredblue}
\end{center}
\end{figure}

Figure \ref{fig:figavgprofiles} presents average spectra of \cii\,  (black) and \oi\, (red)  over the pixels enclosed by the polygon in Figures \ref{fig:figcii} and \ref{fig:figoi}.  The vertical scale shows the T$_{MB}$ brightness temperature in Kelvin. The solid green curve is a Gaussian fit to the central portion of the \cii\, spectrum, extended as the dashed green line. The vertical dashed lines indicate the Blue, Central, and Red emission velocity ranges used in Figure \ref{fig:ciiredblue}.  Figure \ref{fig:ciiredblue} shows maps of the \cii\, intensity integrated over the Blue (blue contours), Central (grayscale), and Red (red controus) velocity ranges from Figure \ref{fig:figavgprofiles}.

\section{PDR Structure: An Internally Ionized Bubble}
PDRs are expected to exhibit a stratified structure, with a transition between fully ionized gas, partially ionized (photodissociated) gas, and fully molecular gas as a function of depth into a dusty, molecular cloud away from the ionizing ultraviolet radiation.  Our observations probe each of these zones.  Ionized gas is traced by radio free-free continuum emission, photodissociated gas by \cii\, and \oi\, line emission, and molecular gas by \nht\, or other molecular line emission.  Since the ionizing OB stars lie to the interior of the Nessie Bubble, theory predicts radio continuum on the interior of the Bubble, followed by a layer of \cii\, and \oi\, emission further toward the exterior, and, finally, \nht\ or other molecular line emission on the outside.

The observed gas distribution generally agrees with this prediction.  The \cii\, (Fig. \ref{fig:figcii}) and \oi\, (Fig. \ref{fig:figoi}) emission roughly has the shape of a question mark, and corresponds to the bright mid-IR emission on the western rim of the Bubble.  This emission delineates the photodissociated gas, presumably powered by the cluster of OB star candidates marked by the red star symbols in Fig. \ref{fig:figcii}.  

Figure \ref{fig:figmapall} shows that the ionized gas, traced by the extended 24 GHz radio continuum, is indeed found toward the Bubble's interior, closest to the stellar cluster.  In addition, the radio continuum also reveals a bright radio continuum point source in the northeast, Position 3, at the tip of the ``question mark,'' coincident with two closely spaced OB star candidates
\citep{Messineo2018}.  This source is likely an embedded compact \hii\, region internally ionized by one or both of the candidate OB stars at Position 3.  Thus, overall, ionized gas lies closest to the interior of the Bubble, a result confirmed by the identical morphology of various recombination lines to that of the radio continuum in the ATCA data.  

In an internally illuminated PDR, the molecular gas should be located toward the Bubble's exterior. In the northern portion of the ``question mark'' the \nht\, (1,1) emission lies at the exterior of the Bubble as expected.  Since this emission is well beyond the 2$'$ FWHM of the ATCA primary beam gain pattern, the emission is actually stronger than portrayed.  (No correction has been applied to the \nht\, images for the primary beam pattern.)  At the base of the crook of the ``question mark,'' \nht\, (1,1) emission reveals a bright, compact source and a filamentary extension from this source to the west.  The compact source is coincident with the luminous FIR/submm YSO AGAL337.916-00.477 (Position 7 in  Fig. \ref{fig:figcii}).  The \nht\, emission here is tracing the molecular gas associated with this embedded YSO.  The filamentary \nht\, emission extending to the west coincides with mid-IR extinction associated with the Nessie IRDC filament.  This filamentary \nht\, emission, therefore, traces gas in the Nessie IRDC.  The position of the luminous YSO lies precisely at the intersection of the Bubble with the IRDC filament.  This location indicates that the interaction between the expanding Bubble and the Nessie filament may have triggered star formation (see Section 6).  Although \nht\, emission indicates molecular gas on the exterior of the Bubble toward the northern portion of the ``question mark,'' it is not detected over the entire ``question mark.''  However, \hcopnt, HCN, \nthpnt, and HNC $1-0$ emission mapped with the ATNF Mopra telescope is in fact detected on the exterior of the Bubble along the entire extent of the ``question mark''  \citep{Jackson2010}.  The lack of \nht\, emission can be explained as due to more extended emission filtered out by the interferometer, or a lack of sufficient \nht\, optical depth or excitation temperature for detection with ATCA.  Thus, the ionized, photodissociated, and molecular gas distributions all conform to the expectations for an internally ionized \hii\, bubble.

\section{\cii\, and \oi\, Absorption and Self-Absorption}
\begin{figure}
\begin{center}
\includegraphics[scale=0.75, angle=0]{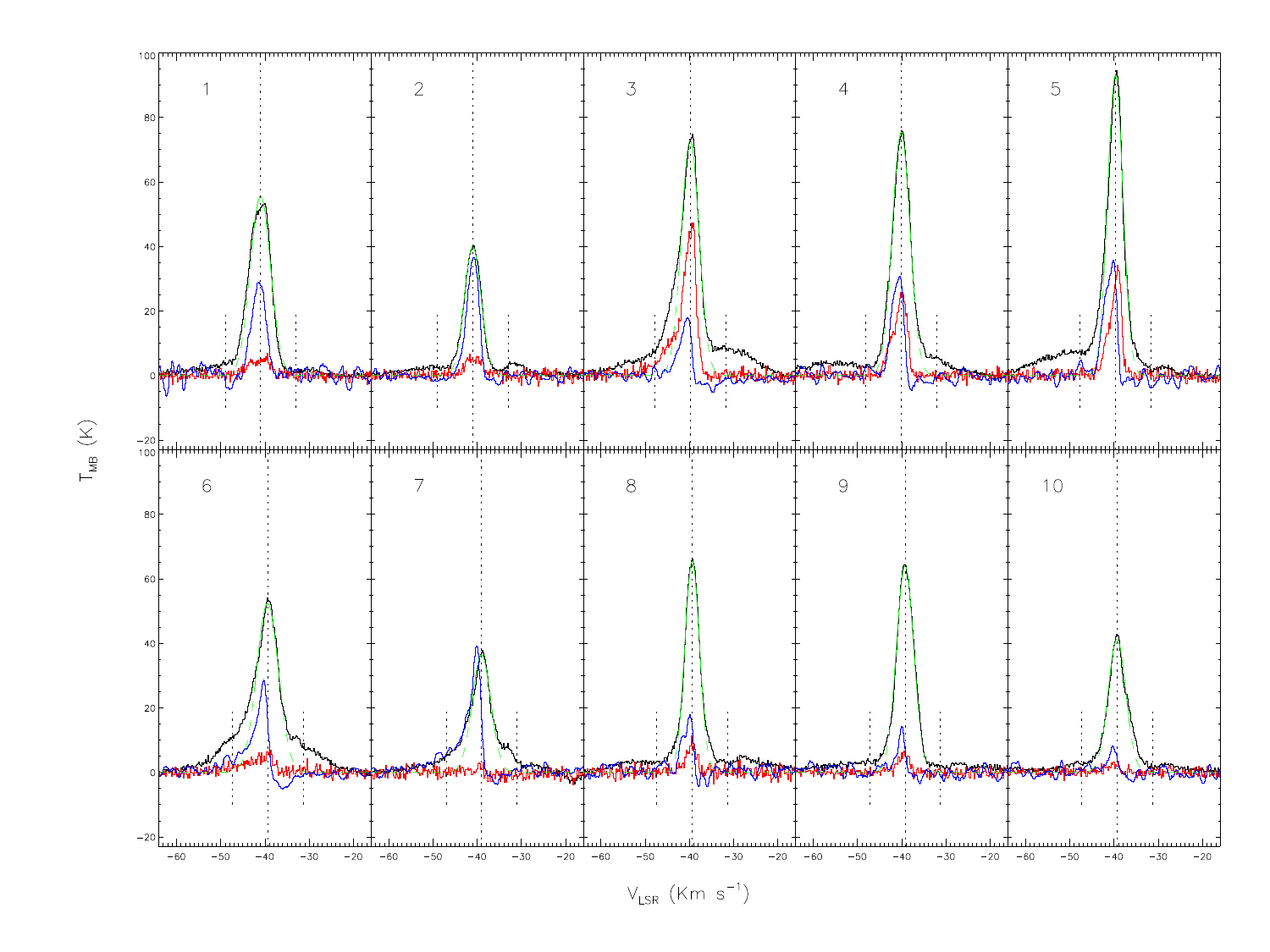} 
\caption{ Spectra of \cii\,  (black trace) and \oi\, (red trace)  at the numbered spots indicated in Figure \ref{fig:figcii}, with green indicating the Gaussian fit to the \cii\ spectra and with axes as in Figure \ref{fig:figavgprofiles}.  The {\bf spectra} for \hcop, measured with the Mopra telescope, multiplied by a factor of twenty, and plotted on the \tastar\, scale, {\bf are} shown in blue.}
\label{fig:figposprofiles}
\end{center}
\end{figure}

Throughout the Nessie Bubble, the \oi\, line shows asymmetric line shapes best explained as arising from deep self-absorption due to lower excitation foreground gas absorbing background PDR line emission.  Although the \oi\, profiles vary considerably at different positions across the cloud, their shapes are typically asymmetric, with more flux on the blueshifted side, or flat-topped (see Fig. \ref{fig:figposprofiles}).  Indeed, the profile of the  \oi\, line spatially integrated over the entire region clearly shows a distinctly asymmetric shape, with stronger emission to the blueshifted side of the profile.  The spatially averaged \cii\, profile, on the other hand, shows a more symmetric shape (see Fig. \ref{fig:figavgprofiles}).  The \hcop\, profiles (Fig. \ref{fig:figposprofiles}) from the Mopra telescope are also asymmetric and typically stronger on the blueshifted side.  Toward several positions, the \hcop\, profiles show an inverse P Cygni profile, with blueshifted emission and redshifted absorption.  Such profiles can arise from an internally heated, collapsing cloud, and may indicate local collapse of the star-forming cores.  The difference between the \oi\, and \cii\, profiles suggest that the \oi\, line is more deeply self-absorbed, i.e., the optical depth of the absorbing gas is higher for \oi\, than that for \cii.   Moreover, toward the bright continuum source AGAL337.916-00.477, at the location where the Bubble is impacting the Nessie filament, both \cii\, and \oi\, show absorption features against the continuum (see Fig. \ref{fig:figcontprofiles}) at \vlsr\, = $-18$ \kms.  In this section, we discuss the idea that \cii\, and \oi\, absorption and self-absorption features are due to subthermally excited ionized carbon and neutral oxygen in the photodissociated diffuse outer regions of giant molecular clouds externally illuminated by a standard interstellar radiation field.

\subsection{{\rm \cii} and {\rm \oi} Absorption Associated with Molecular Cloud Exteriors}

The \cii\, and \oi\, absorption feature toward the bright continuum source AGAL337.916-00.477 (also called GRS G337.92-00.48 or IRAS 16374-4701) at $-18$ \kms\, matches the velocity of a CO $1-0$ emission line detected in the ThrUMMS survey (see Fig. \ref{fig:figcontprofiles}). The continuum emission arises from an embedded, compact, luminous YSO.  The most likely location of this continuum source is within the Nessie IRDC itself, since it is coincident with bright, compact peaks in molecular line emission at the same velocity as the Bubble and the IRDC ($\sim -40$ \kms; see, e.g., the NH$_3$ image in Figure \ref{fig:figmapall}.)  An additional faint \cii\, absorption feature at +3 \kms\, is also marginally detected.  Its association with an absorption feature in \hi\, and an emission feature in CO indicates it is likely to be real.  Here we concentrate our analysis on the $-18$ \kms\, feature, but the same analysis applies equally well to the +3 \kms\, feature.

The distances to both the \cii\, emission and the \cii\, and \oi\, absorption features can be estimated from their velocities using the BeSSeL distance calculator \citep{Reid2019}.  The \cii\, peak velocity of -39.6 \kms\,  associated with the Nessie Bubble and IRDC results in a distance of 2.7 $\pm$ 0.3 kpc \citep{Reid2019}.   This corresponds to the position of the Scutum-Centaurus-OSC arm in the model of the Milky Way of \cite{Reid2019},  extrapolated into the fourth quadrant of the Galaxy.  On the other hand, the $-18$ \kms\, velocity of the absorption features yields a most probable distance (P = 0.7) of 1.25 $\pm$ 0.21 kpc and an association with the extrapolated Sagittarius-Carina arm in the  \cite{Reid2019} model.  
Thus, the $-18$ \kms\, absorption feature is most likely associated with a foreground CO-emitting molecular cloud in an intervening spiral arm and unrelated to the continuum source.  Since the foreground cloud is devoid of any obvious star formation, the C$^+$ and O$^0$ responsible for the absorption is very unlikely to originate in an embedded PDR associated with star-formation.  Instead, the absorbing gas at $-18$ \kms\, must be located elsewhere in the cloud and produced by other means.

We hypothesize that the absorbing gas arises in the periphery of a foreground molecular cloud externally illuminated by a standard insterstellar radiation field.  Any absorption feature must satisfy two conditions: (1) an excitation temperature smaller than the background continuum brightness temperature ($T_{ex} < T_{cont}$) and (2) sufficient optical depth to produce a detectable absorption signal ($\tau_\nu \gtrsim 1$).  Below we show that each of these criteria is met for both the \cii\, 158 \um\, and the \oi\, 63 \um\, lines in the exteriors of all molecular or translucent clouds.

Consider a molecular cloud with no internal source of ultraviolet radiation.  Such a cloud is nevertheless still exposed to an external interstellar ultraviolet radiation field. According to PDR theory, a standard interstellar ultraviolet radiation field ($G_0 \sim 1$) impinging on the exterior of a cloud will photoionize carbon and photodissociate oxygen-bearing molecules to a depth of $A_V \sim 2$ mag (e.g., \citealt{Kaufman2006, Pound2008}) in the cloud periphery.  (Estimates of this depth depend on assumptions about composition, chemical reaction rates, the impinging radiation field, and dust properties.)  Thus, every molecular cloud should have a photodissociated outer layer containing both  C$^+$ and O$^0$.  If this outer layer of  C$^+$ and O$^0$ has sufficiently low excitation temperatures and sufficiently large optical depths, \cii\, and \oi\, absorption features will occur.

For typical molecular clouds, \cii\, and \oi\, will naturally have low excitation temperatures ($T_{ex} \sim 20$ K) due to the fact that, as discussed below, typical molecular cloud densities are far below the \cii\, and \oi\, critical densities.  Thus, the excitation is subthermal, and the excitation temperature is much smaller than the gas kintetic temperature due to the rarity of collisional excitation.  Because radiative transitions dominate over collisional transitions when $n << n_{crit}$, C$^+$ and O$^0$ tend to remain in their ground states, and the excitation temperature is closer to the radiation temperature than the kinetic temperature.  Indeed, in their analysis of \cii\, absorption toward several star-forming molecular clouds, \cite{Gerin2015} conclude that the \cii\, absorption lines must arise from subthermally excited \cii\, with typical excitation temperatures $T_{ex} = 20$ K and gas densities of $n \sim 60$ \pcc.  Toward molecular clouds with no internal star-formation, dust continuum emission is insignificant, and the radiation temperature will be close to that of the cosmic microwave background ($T_R = 2.7$ K).  Consequently, externally illuminated molecular clouds should have low excitation temperatures in the \cii\, and \oi\, lines.

\begin{figure}
\begin{center}
\includegraphics[scale=0.75, angle=0]{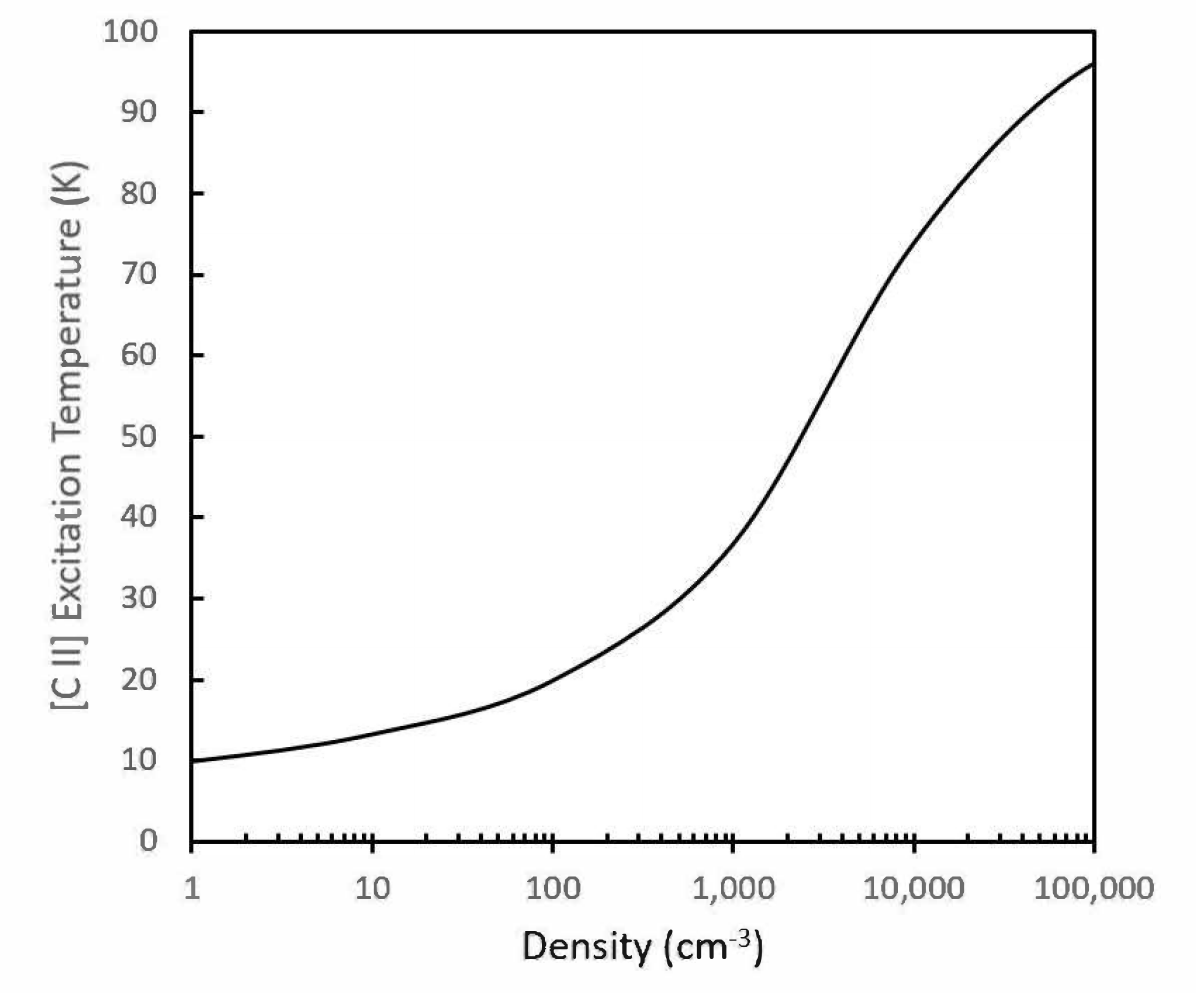} 
\caption{Plot of the \cii\, excitation temperature $T_{ex}$ (vertical axis) versus the particle density (horiziontal axis).  This calculation assumes a critical density for the \cii\, line of 3800 cm$^{-3}$, as appropriate for collisions with atomic hydrogen, and a radiation temperature equal to the microwave background temperature ($T_{R} = 2.7$ K).  Dust continuum radiation from the cloud is optically thin at 158 \um\, and considered to be negligible.  A kinetic temperature of 100 K has been assumed.}
\end{center}
\label{fig:figciiTexvsn}
\end{figure}

First, we demonstrate subthermal excitation in the \cii\, line for typical molecular clouds. For a two-level system such as the [C II] fine-strucure energy levels, the excitation tempaerature $T_{ex}$ is related to the gas kinetic temperature $T_K$ and the radiation temperature $T_R$ by 

$$e^{-h\nu/kT_{ex}} ={{{ n\over{n_{crit}}}e^{-h\nu/kT_K} + {1\over{e^{h\nu/kT_R}-1}}}\over{n\over{n_{crit}}}+(1+{{1}\over{e^{h\nu/kT_R}-1}})}.$$
Here $n_{crit}$ is the critical density, the density at which the downward collisional transition rate equals the radiative transition rate.
The critical density of the \cii\, transition depends on the colliding partner and the kinetic temperature \citep{Goldsmith2012}.  At a kinetic temperature of 100 K, for collisions with molecular hydrogen the critical density is 6100 \cmc, and with atomic hydrogen it is 3800 \cmc.  While the densities of compact star forming regions are often large ($n > 10^4$) \cmc, the typical density of giant molecular clouds traced by CO or $^{13}$CO is $\sim 230$ \cmc\, (e.g., \citealt{RomanDuval2010}). In their peripheries, the densities might be expected to be even smaller.  A plot of the preceding equation, Figure \ref{fig:figciiTexvsn} shows the \cii\, excitation temperature as a function of density assuming that the criticial density is 3800 \pcc, appropriate for 100 K gas composed primarily of atomic hydrogen. (Detailed modeling described below justifies this assumed gas temperature.) The typical densities of giant molecular clouds are far below the critical density.  If the periphery is composed primarily of molecular hydrogen, the critical density is about twice as high, and the excitation temperature would be somewhat lower than in Figure \ref{fig:figciiTexvsn}.  For a system with two different colliding partners, each with a distinct critical density $n_{crit,i}$ and a relative abundance $X_i = n_i/n_{tot}$, the effective critical density is 

$$n_{crit,eff} = {{1}\over{{{X_1}\over{n_{crit,1}}}+{{X_2}\over{n_{crit,2}}}}}.$$
Thus, the effective critical density for \cii\, in a PDR is always between the critical densities for \cii\, colliding with atomic or molecular hydrogen.  Because of the low densities in typical molecular clouds, well below the \cii\, critical density, the \cii\, excitation temperature in their peripheries is subthermal, with characteristic values $T_{ex} \lesssim 20$ K.  For a detailed discussion of critical densities of \cii\, and \oi, see \cite{Goldsmith2012, Goldsmith2021}.

We now consider the \oi\, 63 \um\, line.  Since the critical density of the \oi\, line colliding with molecular hydrogen ($n_{crit} = 5 \times 10^5$ \cmc, \citealt{Kaufman2006}) is larger than that of the \cii\, line, the \oi\, 63 \um\, line in the exteriors of molecular clouds will also be subthermally excited and have similarly small excitations temperatures,  $T_{ex} \lesssim 20$ K. 

The peripheries or ``skins'' of molecular clouds also have sufficient optical depth to produce absorption and self-absorption features for both \cii\, and \oi.   We first consider the \cii\, line.  The optical depth of a two-level system is given by

$$ \tau = {\tau}_0 {{1 - e^{-{h\nu/kT_{ex}}}}\over{1 + {(g_u/g_l)}e^{-{h\nu/kT_{ex}}}}},$$
where
${\tau}_0 = {hB_{lu}N(C^+)/\delta v}.$
Here $B_{lu}$ is the downward Einstein B coefficient and $\delta v$ the linewidth, defined as the reciprocal of the line profile function $\phi(v)$ at line center.  For Gaussian line profiles, the value for $\delta v$ is within 7\% of the value for the FWHM linewidth $\Delta V$. 
For \cii, $\tau_0$ can be approximated by 
$\tau_0 = 7.49 \times 10^{-18} N(C^+) ({\rm cm}^{-2})/\delta v ({\rm km~s}^{-1})$ \citep{Goldsmith2012}.

\begin{figure}
\begin{center}
\includegraphics[scale=0.75, angle=0]{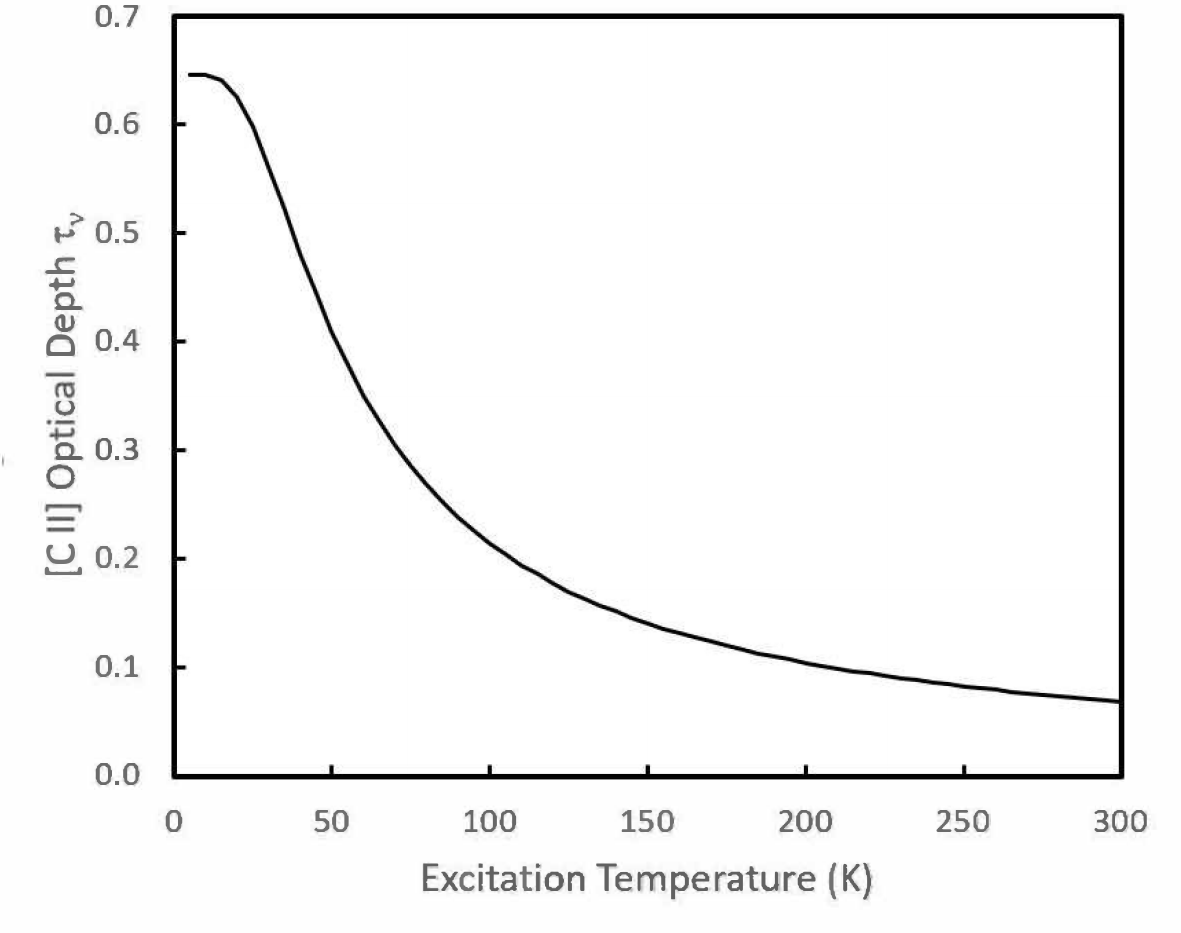} 
\caption{ Plot of the [C II] 157 \um\, optical depth at line center (vertical axis) versus the excitation temperature (horiziontal axis) for a typical giant molecular cloud.  The calculation assumes a visual extinction $A_V = 2$ mag, all of the carbon is ionized, the carbon abundance relative to hydrogen is $1.2 \times 10^{-4}$, the FWHM line width $\Delta V = 5$ \kms, and the relation between visual extinction and column density is 
$N(H) = 1.87 \times 10^{21} A_V$ \cms\, mag$^{-1}$.}
\label{fig:figciitauvsTex}
\end{center}
\end{figure}

To estimate the optical depth of a \cii\, line produced from the outer portions of a molecular cloud, we can make the simplifying assumption that all of the carbon in a zone with $A_V < 2$ mag is ionized. Since PDR models indicate an ionized carbon abundance near unity for $A_V < 2$ mag and a sharp drop to near zero abundance for $A_V > 2$ mag, for rough estimates this assumption is well justified.  Thus, the column density of ionized carbon in the outer portions of a molecular cloud is well-approximated by the hydrogen column density corresponding to $A_V = 2$ mag, multiplied by the cosmic abundance of carbon.  Plugging in the conversion factor $N(H) = 1.87 \times 10^{21} A_V$ \cms\, mag$^{-1}$ \citep{Bohlin1978} and the cosmic abundance of carbon $1.2 \times 10^{-4}$ \citep{WakelamHerbst2008}, we obtain $N(C^+) = 4.5 \times 10^{17}$ \cms\, for the typical column density of ionized carbon in the exterior of a molecular cloud.  Using the equation above, we can calculate the \cii\, optical depth as a function of excitation temperature.  This relation, for an assumed typical $\delta v = 5$ \kms, is plotted in Figure \ref{fig:figciitauvsTex}.  Because this estimate ignores the transition between atomic and molecular hydrogen in the PDR, $N(C^+) $ will be somewhat underestimated, but by less than a factor of 2.

For diffuse gas with $n << n_{crit}$, as expected in the exterior of giant molecular clouds, the excitation temperature is low, $T_{ex}<20$ K, and the \cii\, optical depth is $\tau \sim 0.65$.  Realistically, when background continuum or line radiation passes through a foreground cloud, it should pass through an outer layer twice, once on the back side of the cloud and once on the front side.  Since both absorbing layers will be at the cloud's systemic velocity, the optical depth will be twice that of passage through a single layer, $\tau \sim 1.3$. Indeed, this value agrees well with the optical depth estimated for observed \cii\, self-absorption features (e.g., \citealt{Guevara2020, Kabanovic2022}). Thus, both requirements for producing \cii\, absorption features, either absorption against a background continuum emission or self-absorption against bright background line emission, are met: (1) low excitation temperature, and (2) sufficient optical {\bf depth} with $\tau(\cii) \sim 1$.

Estimating the optical depth of \oi\, is more complicated for two reasons.  First, neutral atomic oxygen is a three-level system, and optical depth estimates  will depend on collisional and radiative transitions into the upper ${^3}P_0$ energy level.  For the low densities in typical giant molecular clouds, however, such excitations have a neglible effect.   Second, the formation and destruction of O$^0$ as a function of $A_V$ is more complicated than that of C$^+$ and models suggest that O$^0$ can exist at significant abundances beyond the PDR and well into the molecular cloud.  The exact extent, however, depends on assumptions about both gas phase and solid phase chemistry.  Below we describe detailed PDR models that demonstrate that the optical depth of the \oi\, 63 \um\, line is also sufficient to produce absorption features.

\subsection{Detailed modeling using the Meudon PDR code}
We use the Meudon PDR models to demonstrate that typical molecular clouds externally illuminated with a standard interstellar radiation field will produce \cii\, and \oi\, 63 \um\, lines that satisfy both of the criteria necessary to produce detectable absorption features, namely, low excitation temperatures and sufficient optical depth.  The Meudon PDR code models the chemistry, thermal balance, and line excitation of clouds exposed to various radiation fields (see \citealt{LePetit2006, Goicoechea2007, GonzalezGarcia2008, LePetit2009} for details).  The most recent Meudon models provide updates to include surface chemistry on dust grains, which has an important effect on the oxygen chemistry due to the formation of water ice in grain mantles.

We have modeled clouds exposed to a standard interstellar radiation field described by \cite{Mathis1983}, which closely approximates a standard Habing field with $G_0 = 1$, but also includes a far-infrared component due to dust.  Both the front side and the back side of the cloud are exposed to this standard field, and thus all models are symmetric about the midpoint.  The models are isobaric, with constant pressures throughout the cloud of $P/k = nT = 10^4$ \pcc\, K.  Characteristic turbulent velocity dispersions are considered to be uniform with a velocity dispersion $\sigma_V = 1$ \kms, equivalent to FWHM $\Delta V = 2.35$   \kms.  We consider three clouds: a ``translucent cloud'' with a total visual extinction $A_V = 1.5$ mag, a ``barely molecular cloud''  with $A_V = 5$ mag, and a ``very molecular cloud'' with $A_V = 20$ mag.  Since the cloud is illuminated from both sides, the visual extinction to the cloud center is half of these total values.  These plane-parallel slab models are meant to explore parameter space and to investigate the excitation temperatures and optical depths more realistically than the simple estimates above. 

\begin{figure}
\begin{center}
\includegraphics[scale=0.33, angle=0]{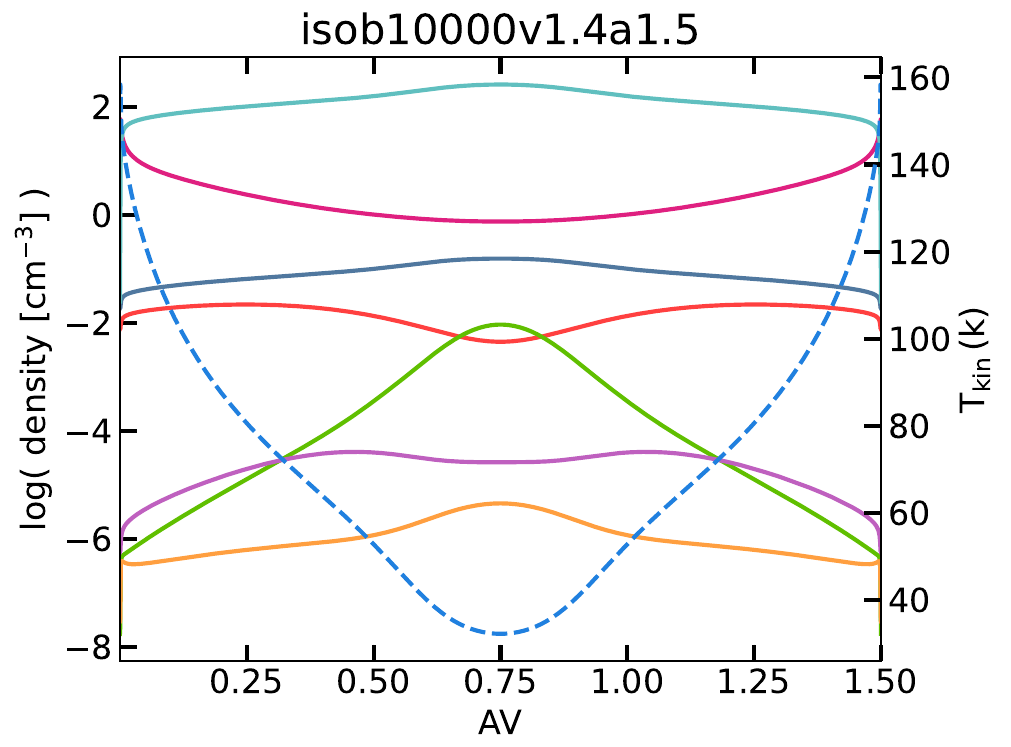} 
\includegraphics[scale=0.33, angle=0]{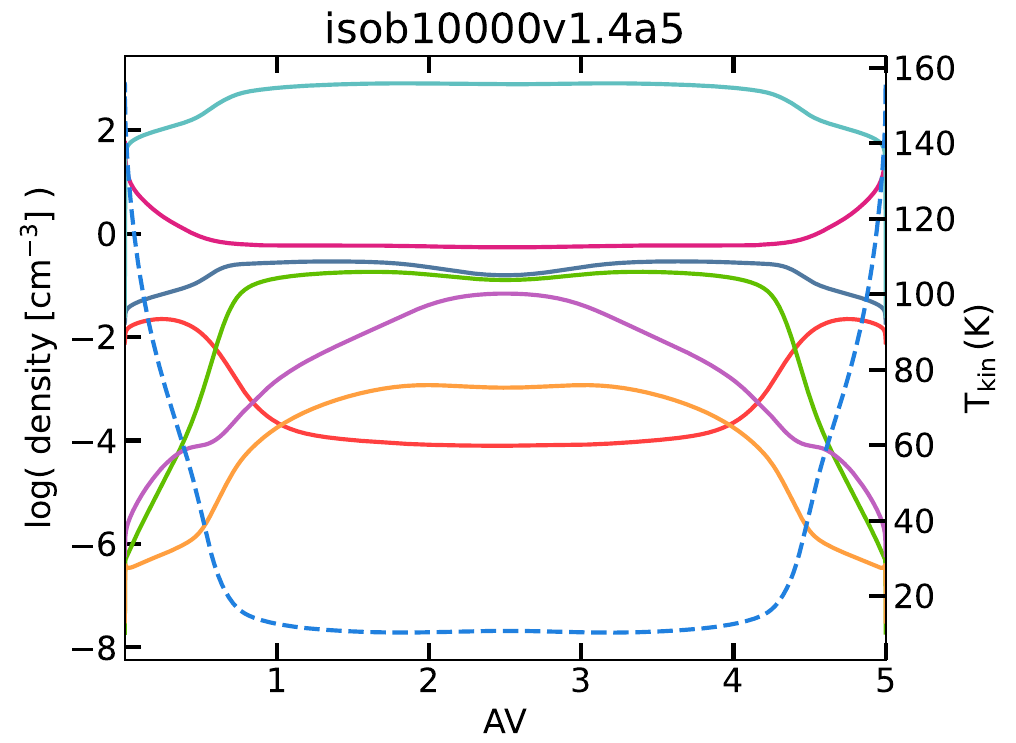} 
\includegraphics[scale=0.33, angle=0]{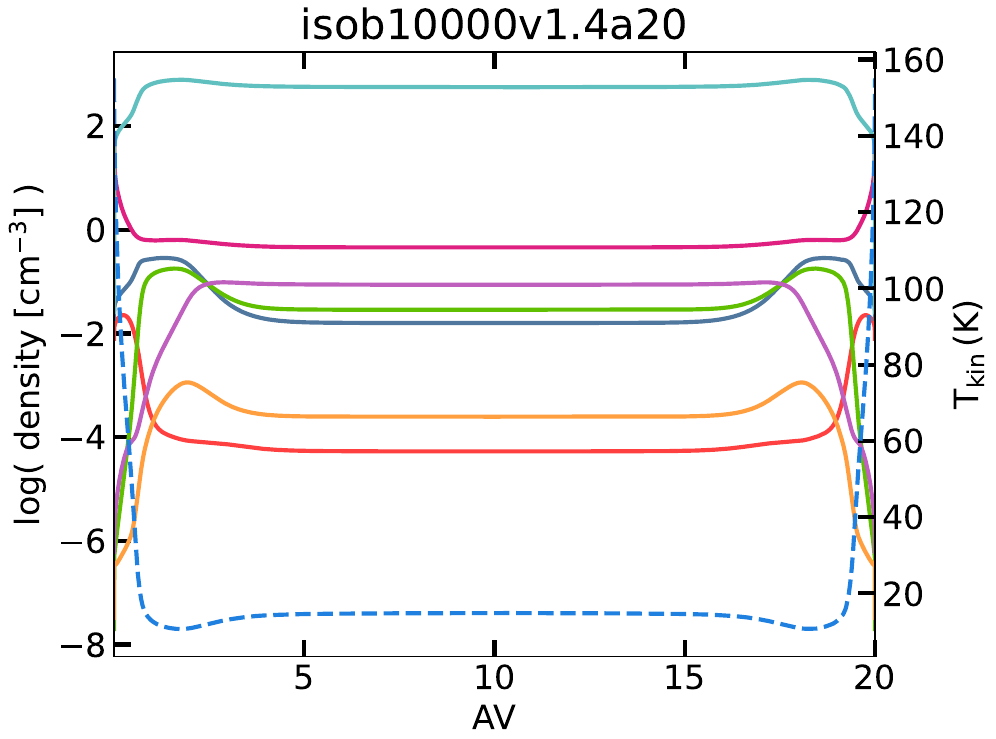} 
\includegraphics[scale=1, angle=0]{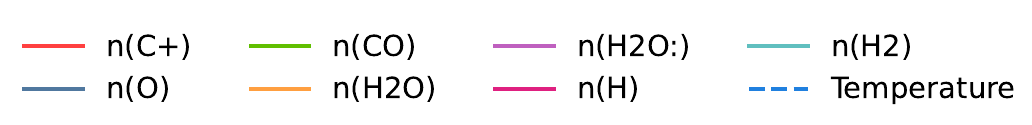}
\caption{Meudon PDR model plots of the volume densities of selected species and the kinetic temperature for isobaric models with $P = nT/k = 10^4$ \pcc\, K.  The densities of C$^+$, O$^0$, CO, H$_2$, H, H$_2$O in the gas phase (gold line, denoted by H$_2$O), and H$_2$O ice (purple line, denoted by H$_2$O:) are plotted as a function of visual extinction.  The kinetic temperature is also plotted as a dashed line.  Three cloud models are shown: (left) ``translucent'' ($A_V = 1.5$ mag), (middle) ``barely molecular'' ($A_V = 5$ mag), and ``very molecular'' ($A_V = 20$ mag).  The clouds are illuminated from both sides by a $G_0 = 1$ radiation field. The velocity dispersion for each cloud is assumed to be $\sigma_V = 1$ \kms.}
\label{fig:Meudon-chemistry}
\end{center}
\end{figure}

Figure \ref{fig:Meudon-chemistry} shows the chemical structure of these three clouds for C$^+$, O$^0$, CO, H, H$_2$, and H$_2$O both in the gas (vapor) phase and the solid (ice) phase.  As expected, C$^+$ is largely confined to the outskirts of the cloud with $A_V < 1$ mag.  (This value is somewhat smaller than the canonical $A_V = 2$ mag extinction for a PDR due to the detailed assumptions used in various models.)  On the other hand, the O$^0$ abundance peaks deeper inside the cloud, between $A_V \sim 1$ to 2 mag.  For larger visual extinctions, oxygen primarily forms water ice in dust mantles, and the O$^0$ abundance becomes small.  Indeed, in the  interior of the``very molecular'' cloud, for $A_V > 2$ mag, the water ice abundance significantly exceeds that of both O$^0$ by factors of a few and also of water vapor by factors of $\sim 100$. Thus, for clouds of small to moderate total extinction, the models suggest that absorption in the 63 \um\, \oi\, line arises mainly in the outer layers of clouds with $A_V < 2$ mag, with a smaller contribution from the interiors of molecular clouds in the absence of internal PDRs.  For clouds with very large column densities, however, the contribution from atomic oxygen in the interior molecular region can contribute significantly to the total optical depth \citep{Goldsmith2021}.

\begin{figure}
\begin{center}
\includegraphics[scale=0.39, angle=0]{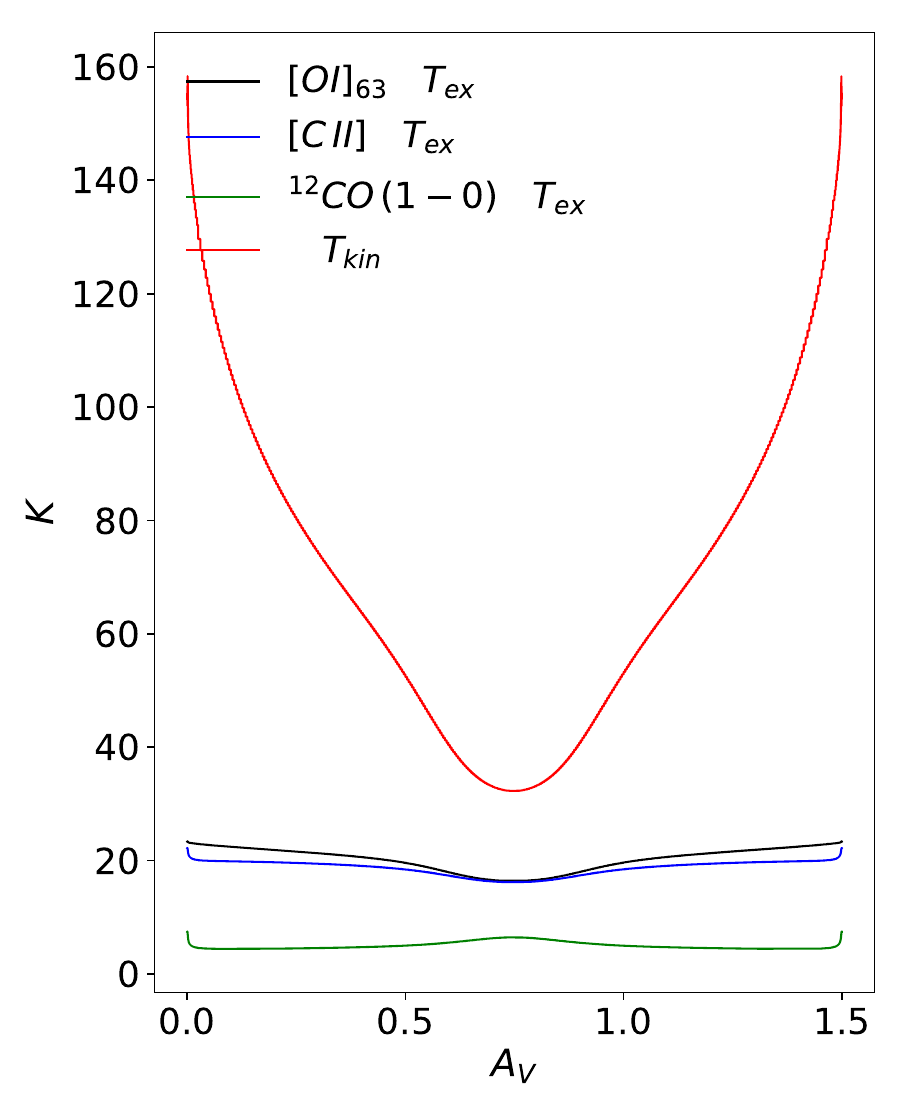} 
\includegraphics[scale=0.39, angle=0]{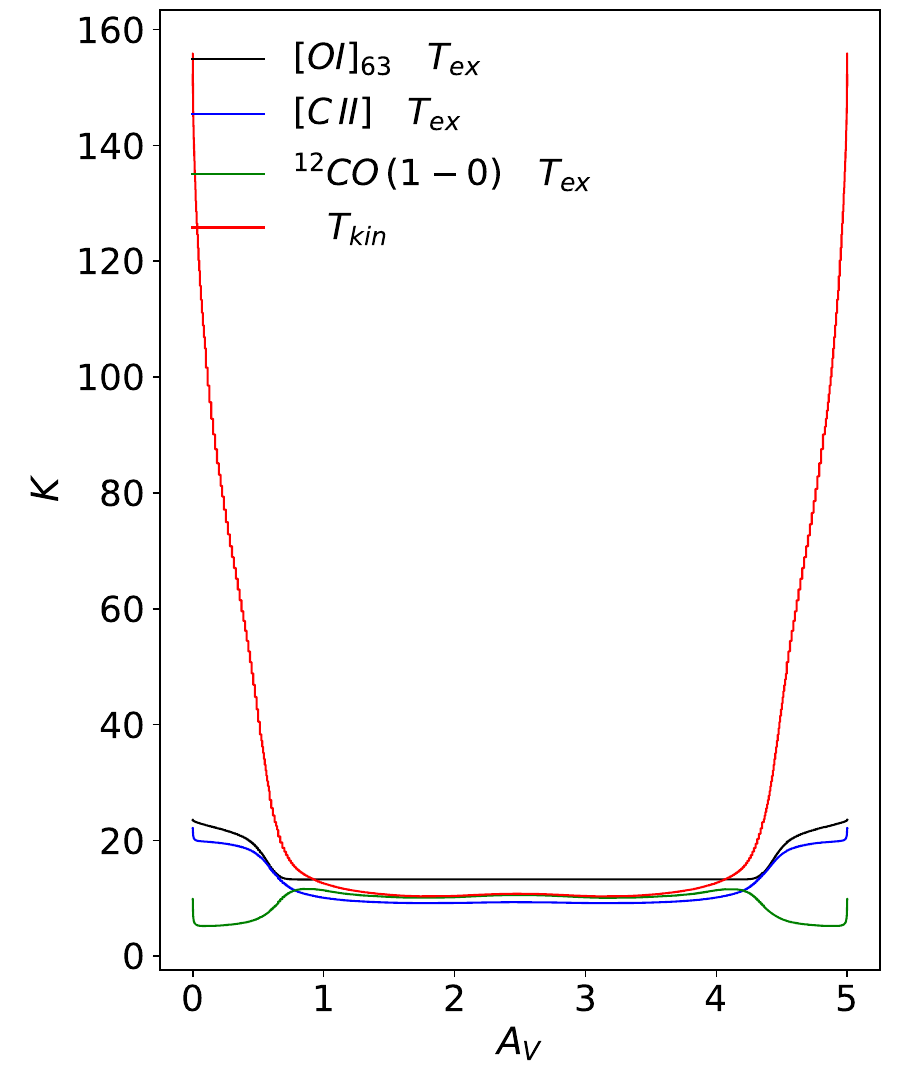}
\includegraphics[scale=0.39, angle=0]{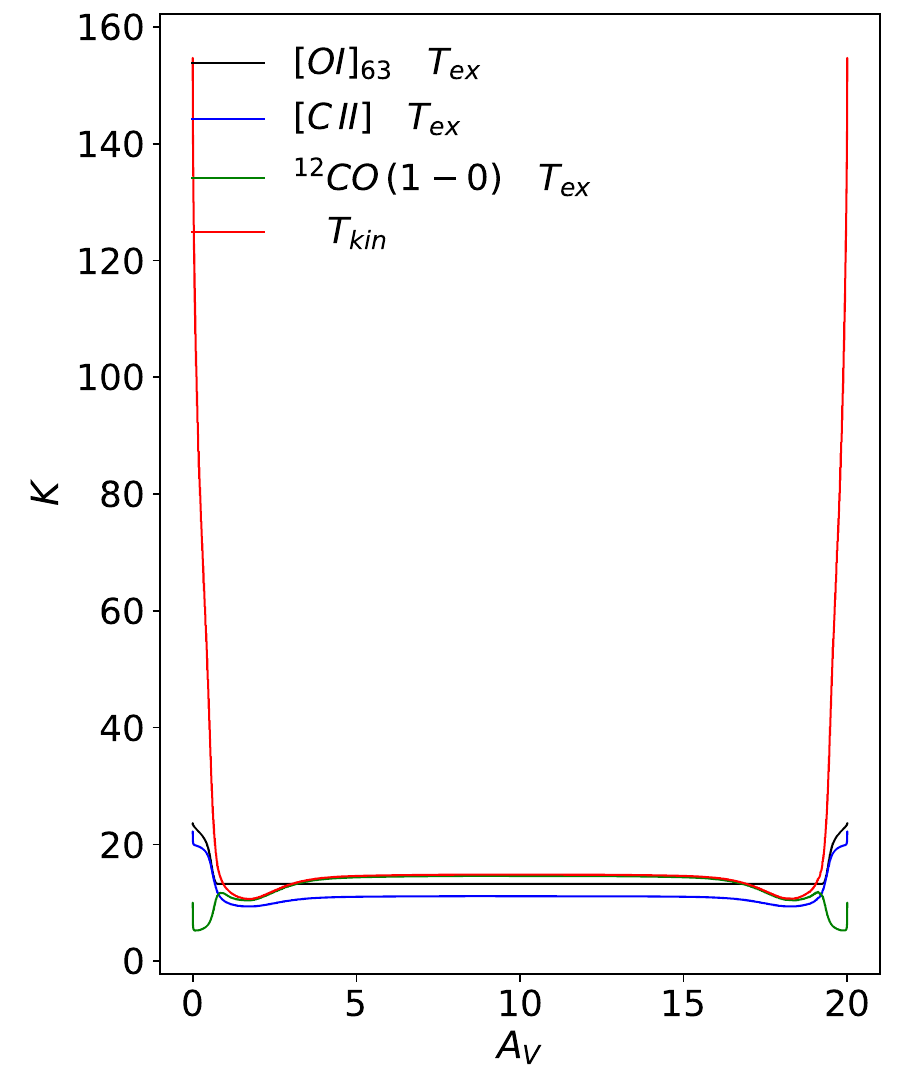}
\caption{Meudon PDR model plots of the kinetic temperature (red), and the excitation temperatures of \oi\, (black), \cii\, (blue) and CO 1-0 (green), as a function of visual extinction into the cloud. Three cloud models are shown: (left) ``translucent'' ($A_V = 1.5$ mag), (middle) ``barely molecular'' ($A_V = 5$ mag), and ``very molecular'' ($A_V = 20$ mag).  The clouds are illuminated from both sides by a $G_0 = 1$ radiation field.  The velocity dispersion for each cloud is assumed to be $\sigma_V = 1$ \kms.  For all three clouds typical excitation temperatures are $\sim 20$ K, independent of the cloud's total visual extinction.}
\label{fig:Meudon-Tex}
\end{center}
\end{figure}

The Meudon models for the representative clouds can be used to estimate the excitation temperatures and the total optical depths for the \cii, \oi\, 63 \um, and CO $1-0$ lines.  Figure \ref{fig:Meudon-Tex} shows the kinetic temperature and the \cii, \oi, and CO excitation temperatures as a function of visual extinction into the cloud.  The detailed modeling confirms the above suggestion that in the cloud exteriors both the \cii\, and \oi\, lines are subthermally excited, with $T_{ex} \sim 20$ K.  Deeper into the ``very molecular'' cloud interiors the actual kinetic temperature approaches 15 K, and for this choice of pressure, the interior densities grow higher and the lines are closer to thermalization.  The overall effect is that excitation temperatures for both \cii\, and \oi\, remain close to 20 K throughout the cloud.  Actual estimates of the excitation temperatures of \cii\, absorption features associated with clouds in the foreground of bright star-forming regions confirm this prediction, with characteristic values of $T_{ex} = 20$ K \citep{Gerin2015}.

\begin{figure}
\begin{center}
\includegraphics[scale=0.35, angle=0]{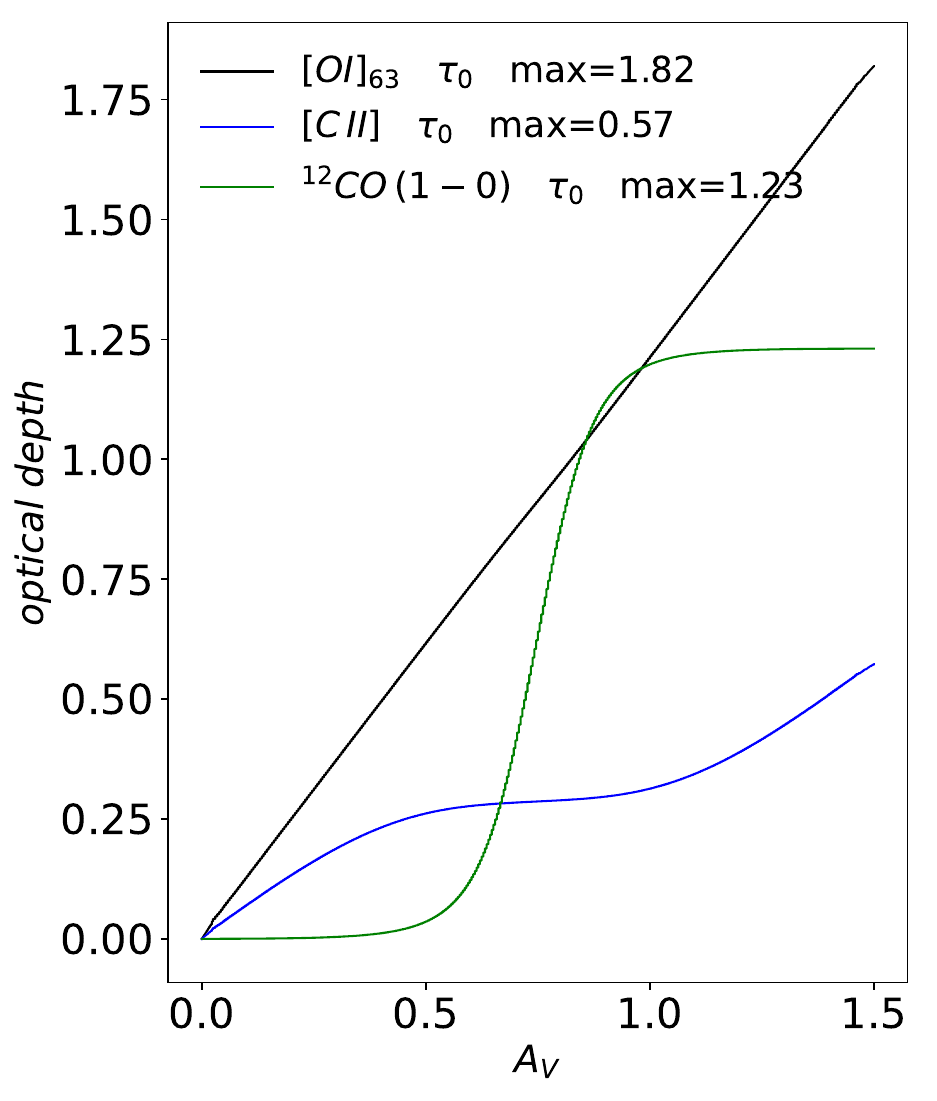}
\includegraphics[scale=0.35, angle=0]{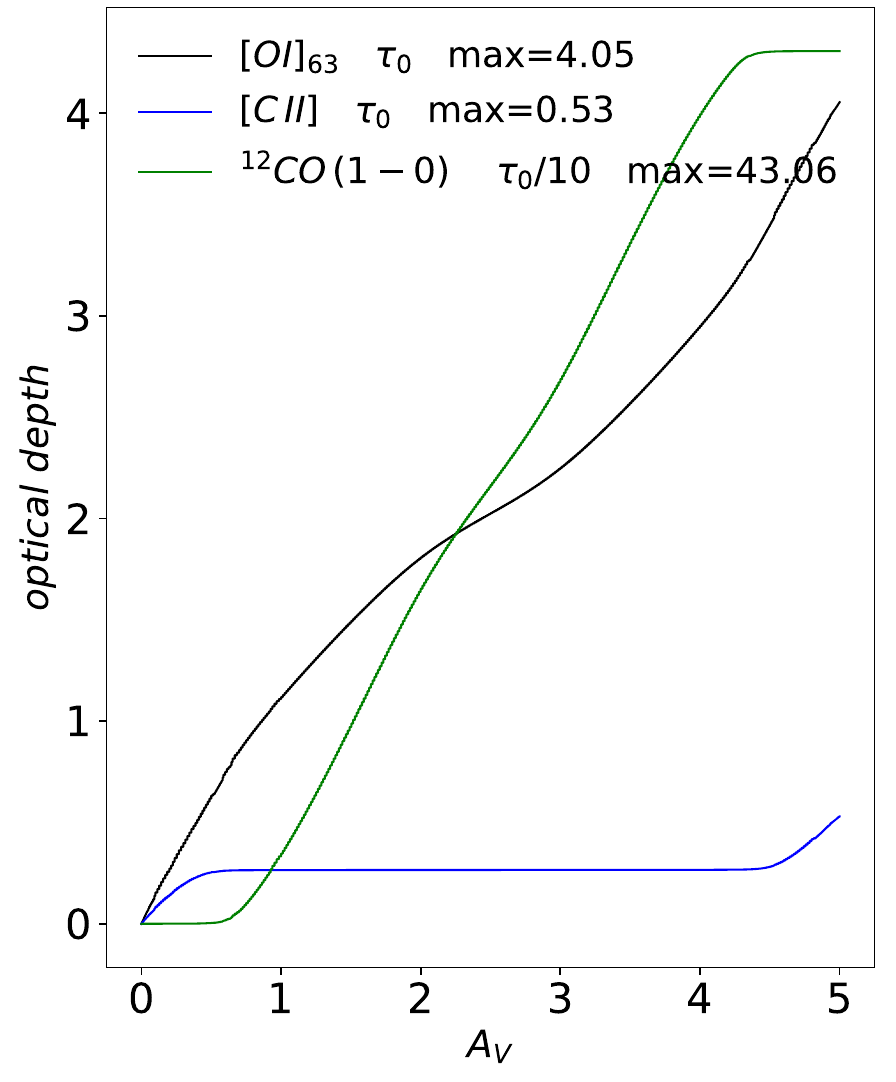}
\includegraphics[scale=0.35, angle=0]{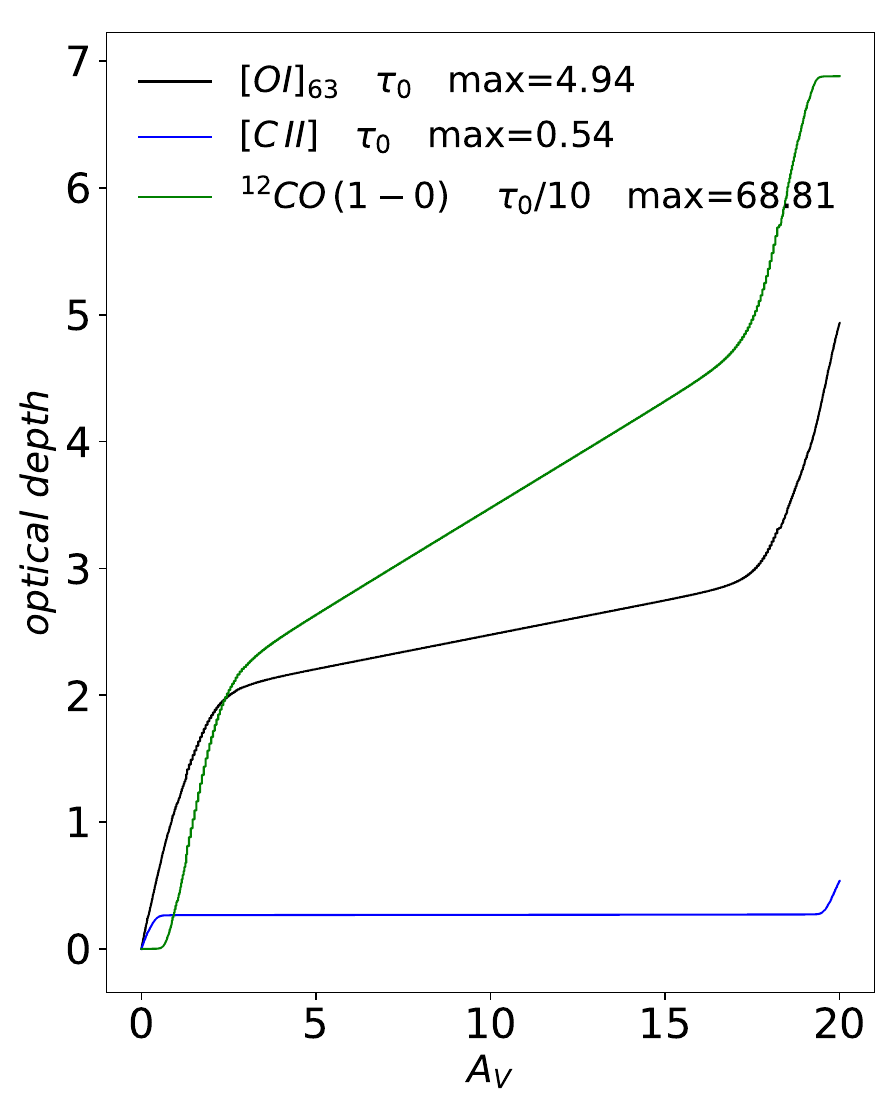}
\caption{Meudon PDR model plots of the integrated optical depth of \oi\, (black), \cii\, (blue) and CO 1-0 (green), as a function of visual extinction into the cloud.  These optical depths are integrated from the front edge to the back edge of the cloud, and thus pass through the photodissociated ``skin'' twice.  Three cloud models are shown: (left) ``translucent'' ($A_V = 1.5$ mag), (middle) ``barely molecular'' ($A_V = 5$ mag), and ``very molecular'' ($A_V = 20$).  The clouds are illuminated on both sides by a $G_0 = 1$ radiation field.  The velocity dispersion for each cloud is assumed to be $\sigma_V = 1$ \kms.  For all three clouds the optical depths of \cii\, exceeds 0.5 and that of \oi\, exceeds 1.8.  In the middle and right panels, the CO optical depth has been divided by 10.}
\label{fig:Meudon-tau}
\end{center}
\end{figure}

Figure \ref{fig:Meudon-tau} displays the integrated optical depth at line center through the cloud as a function of visual extinction for \cii, \oi, and CO $1-0$.  For all three clouds the \cii\, optical depth is $\tau \sim 0.5$.  This equality demonstrates that the \cii\, optical depth arises solely from the exterior layer of the cloud, with no significant contribution from the interior beyond $A_V > 1$ mag.   The \oi\, total optical depth climbs from 1.8 for the ``translucent cloud'' to 4.1 for the ``barely molecular'' cloud, and on to 4.9 for the ``very molecular'' cloud.  This result suggests that the bulk of the \oi\, optical depth arises between $A_V = 1$ to 2 mag.  Adding another 15 magnitudes of visual extinction to the $A_V = 5$ mag cloud does not significantly increase the \oi\, optical depth.  Conversely, the CO $1-0$ optical depth increases significantly and steadily as the cloud's total visual extinction grows.  

\begin{figure}
\begin{center}
\includegraphics[scale=0.7, angle=0]{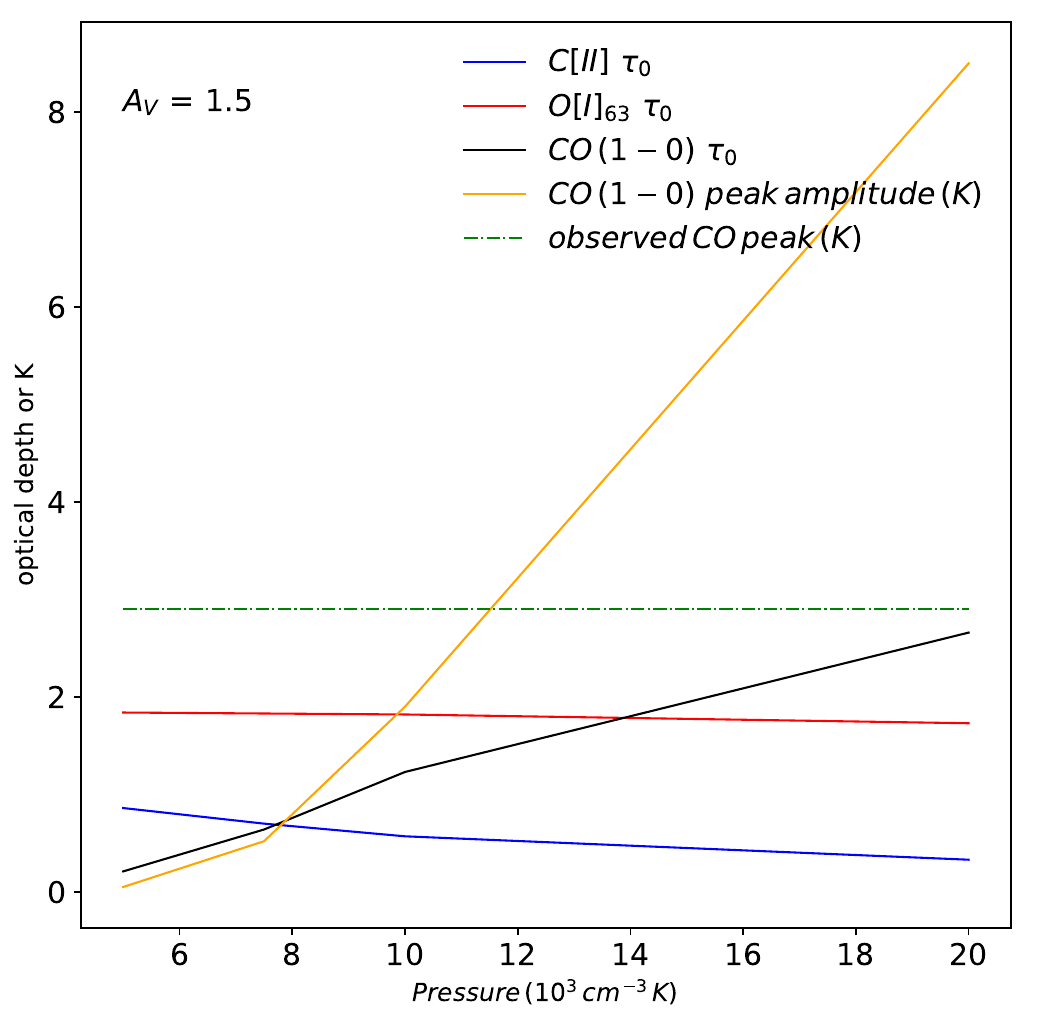} 
\caption{A comparison of the isobaric Meudon model predictions for a foreground cloud with $A_V = 1.5$ and $\sigma_V = 1$ \kms\, as a function of pressure $nT$.   The plot displays the optical depth at line center of \cii\, (blue), \oi\, (red), and CO $1-0$ (black).  Also shown are the predicted peak amplitude of the continuum subtracted CO line (gold) and the actual observed peak amplitude of the CO line (dashed green line).  Both the \cii\, and \oi\, optical depths are slightly higher for lower pressures.
}
\label{fig:pressplots}
\end{center}
\end{figure}

The choice of pressure will change the overall values of optical depth.  We have chosen isobaric models with a constant pressure $nT = 10^4$ \pcc\, K throughout the cloud.  The model results for different pressures is shown in Figure \ref{fig:pressplots}. For the ``translucent cloud'' models with $A_V=1.5$ mag, the \cii\, and \oi\, optical depths are higher for lower pressures, and the change in the \cii\, optical depth is more significant.  For example, in the models of ``barely molecular'' clouds with total $A_V = 1.5$ mag, for $nT = 5,000$ \pcc\, K, the \cii\, optical depth grows to 0.86, compared to a value of 0.57 for $nT = 10,000$ \pcc\, K.

\begin{figure}
\begin{center}
\includegraphics{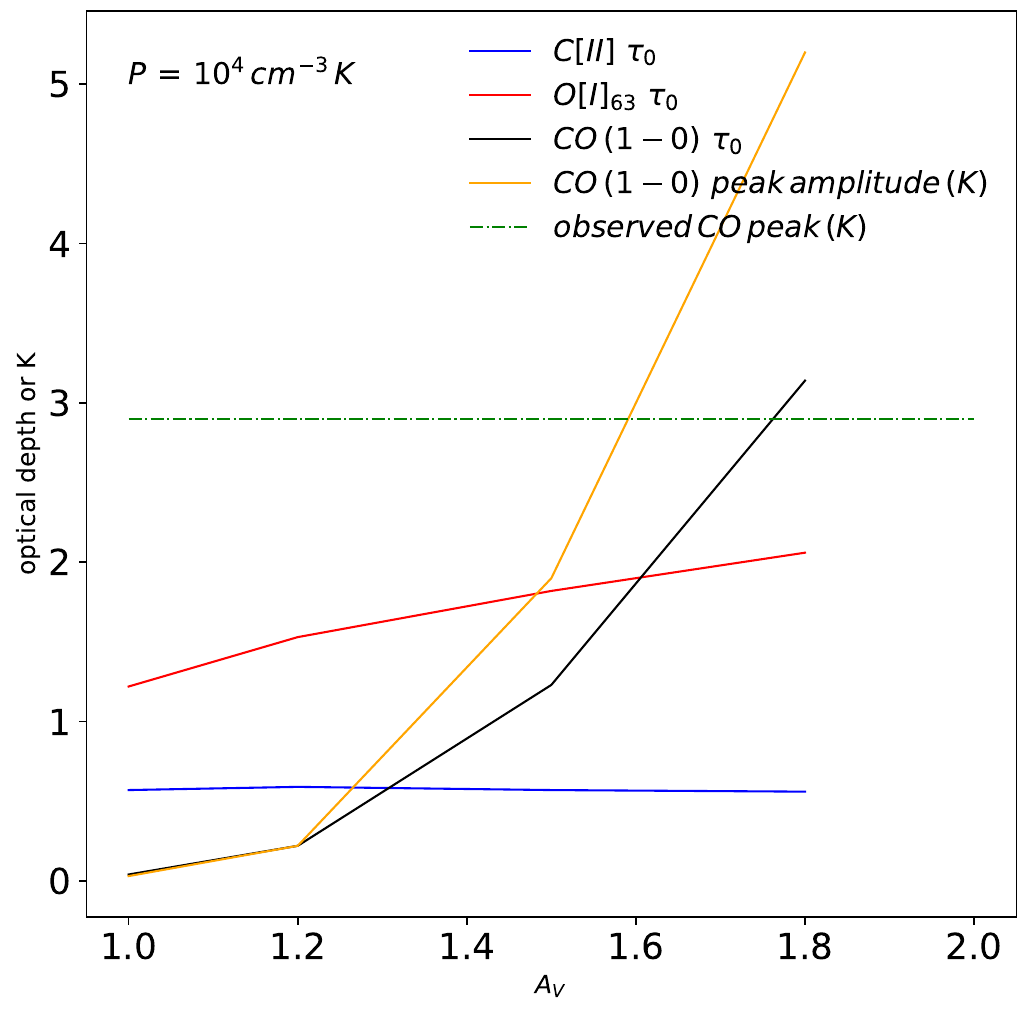} 
\caption{ Meudon model values for the line-center optical depths  for the \cii\, (blue), \oi\, (red), and CO(1-0) (black) lines as a function of  the slab thickness in $A_V$.  The models were isobaric with $nT = 10^4$ K  \cmc\, and the slabs were illuminated from each side by a standard interstellar radiation field with intensity of 1.0 in Mathis units.  The velocity dispersion for each cloud is assumed to be $\sigma_V = 1$ \kms. Also shown, in orange, is the peak amplitude (K) above the continuum of the calculated CO (1-0) emission.  \cii\, and \oi\, emission were negligible in these models.  The horizontal dashed green line shows the peak amplitude of the gaussian fit to the observed continuum-subtracted CO (1-0) emission.}
\label{fig:avplots}
\end{center}
\end{figure}

\begin{figure}
\begin{center}
\includegraphics[scale=0.5, angle=0]{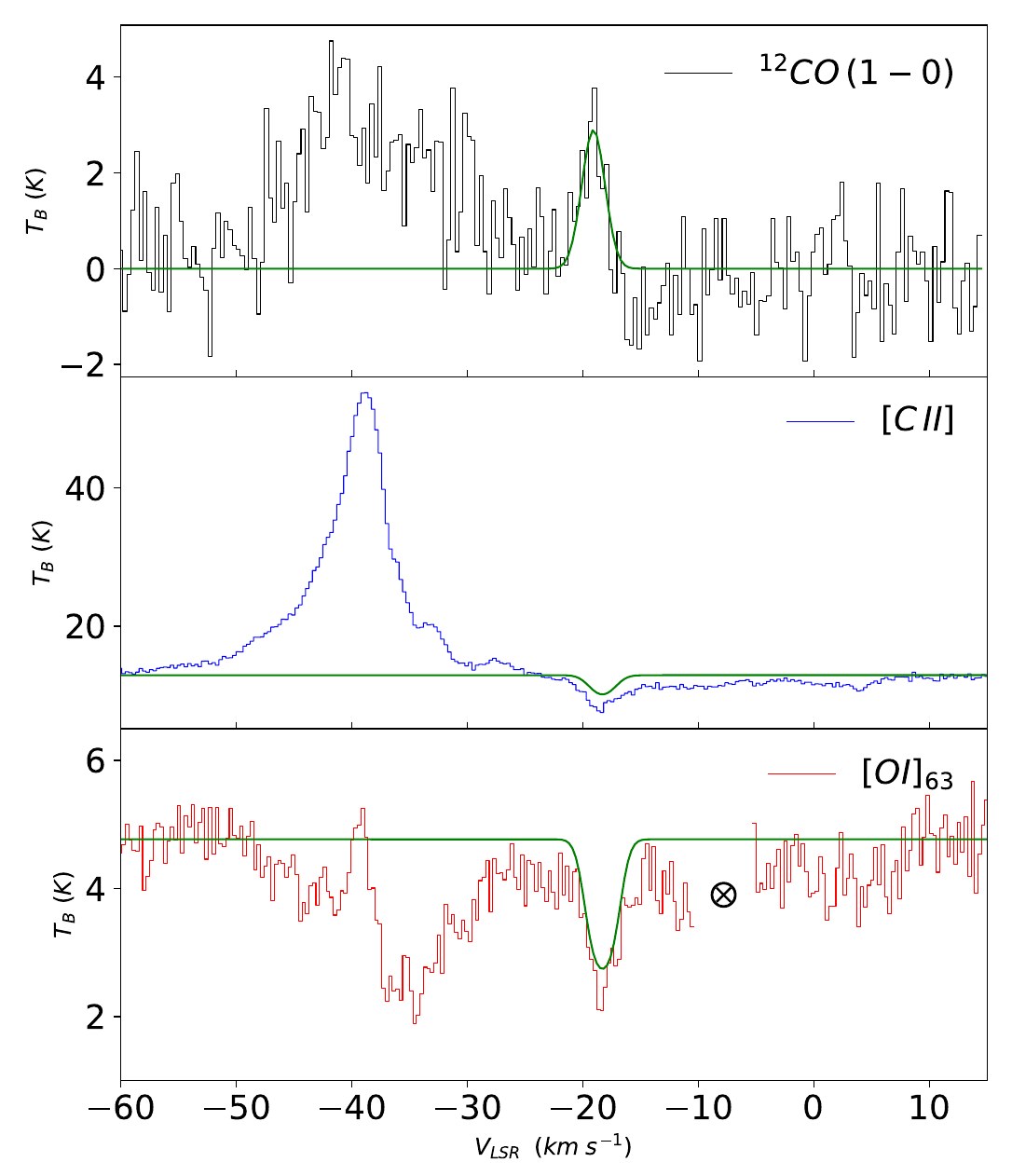} 
\caption{A comparison of the Meudon model predictions for a foreground cloud with $A_V = 1.57$ with the observed data towards AGAL337.916-0.477.  Top: continuum-subtracted CO (1-0) emission (THRUMMS); the green line is a gaussian fit with peak amplitude 2.9 K, center velocity of -18.3 \kms, and rms velocity dispersion of 1.0 \kms.  Middle: \cii\, emission with continuum.  Since upGREAT was a double-sideband instrument, the constant level is twice the actual continuum intensity.  
The green line is a plot of the expression 
$T_B = 2T_C - T_C(1-e^{-\tau_0})e^{-({{V-V_0}\over{\delta V}})^2}$
with $T_C=6.4$ K, $V_0 = -18.3$ \kms, $\delta V = 1.4$ \kms, and $\tau_{0} =0.8$.   Bottom: \oi\, emission with continuum.  A telluric feature around -8 \kms\, has been blanked.  The green line is a model calculation as in the Middle figure, with  $T_C=2.4K$, $V_0 = -18.3$ \kms, $\delta V = 1.4$ \kms, and $\tau_{0} =2.6$.    Continuum values for \cii\, and \oi\, are the average of the intensities over the range $10 < V < 60$ \kms.  
The optical depths for \cii\, and \oi\, were taken from the Meudon model curves (see Figure \ref{fig:avplots}) for $A_V = 1.57$, motivated by the observed CO (1-0) peak intensity.
}
\label{fig:specfit}
\end{center}
\end{figure}

We can constrain the properties of the foreground absorbing cloud by matching the observations to the model predictions.  Figure \ref{fig:avplots} shows the predicted values of the \cii, \oi\, and CO optical depths as a function of visual extinction for the foreground absorbing cloud, as well as the predicted and observed CO brightness temperatures.  Here we have chosen a velocity dispersion $\sigma_V$ = 1.0 \kms\, in order to match the observed CO line width. The observed CO brightness temperature is shown as a horizontal dashed green line.  By matching the observed CO brightness with the predicted one, we can constrain the extinction of the foreground cloud to $A_V = 1.6$ mag, close to that of the nominal ``translucent'' cloud. This value for the visual extinction also fixes the predicted values for the \cii\, and \oi\, optical depths {\bf to values of} $\tau_0$ = 0.6 and 1.9 for \cii\, and \oi\, respectively.  Using these parameters predicted by the Meudon models, we can then examine how well they match the data.  Figure \ref{fig:specfit} shows the results. For both CO and \oi, the agreement is very good.  For \cii, the absorption dip is deeper than the model predicts.  However, the presence of significant line wings in the \cii\, spectrum (see Fig. \ref{fig:figposprofiles}) suggests that, in addition to the foreground cloud, additional \cii\, gas likely exists at this velocity, and thus there may be additional absorption due to this gas. Given the simplicity of the homogeneous, slab model, the agreement between the Meudon model and the data is satisfactory, and demonstrates the plausibility of the ``skin'' model for this souce. To summarize, the predictions of the Meudon model for a single, foreground cloud with $A_V = 1.6$ mag and externally illuminated with a standard $G_0$ = 1 insterstellar radiations field match the observed CO, \cii\, and \oi\, data reasonably well for the foreground cloud along the line of sight to AGAL337.916$-$00.477.

The Meudon models demonstrate that the exteriors of molecular clouds are plausible sites of the widely observed absorption features in \cii\, and \oi.  The models suggest that, toward any cloud with a total $A_V \gtrsim 1.5$ mag, in the peripheries or `'skins'' of clouds, both lines have low excitation temperatures ($T_{ex} \lesssim 20$ K) and sufficient optical depths ($\tau_\nu \gtrsim 0.5$) to produce absoption or self-absorption features.  We interpret the \oi\, self-absorption, and the \cii\, and \oi\, absorption features toward the Nessie Bubble as arising from the exteriors of molecular clouds.  In the case of self-absorption, the likely cloud is the Nessie Nebula itself, but in the case of the foreground absorption features at different velocities, the absorption arises from intervening foreground clouds.  Given the generality of this analysis, it is tempting to explain many, and perhaps most, such \cii\, and \oi\, self-absorption and absorption features in the same way.  

Self-absorption of \cii\, and \oi\, emission lines from the PDR associated with a central, embedded star-forming region in a molecular cloud can readily be explained by this scenario.  The line emission from a bright PDR in the the center of a cloud must pass through the periphery of its own parental cloud on the way to the observer.  Since the periphery of any cloud should contain subthermally excited \cii\, and \oi\, both with $\tau \gtrsim 1$, every sufficiently bright PDR should display a \cii\, and \oi\, self-absorption feature at the parental cloud's systemic velocity.  The ubiquity of such self-absorption features supports this prediction.

\cite{Goldsmith2021} have used the Meudon models to perform a similar study to explain the \oi\, self-absorption toward W3.  They model the cloud as having an embedded PDR with $G_0 = 10^5$, a density $n = 10^5$ \cmc, and a total extinction of $A_V = 100$ mag.  They model the line profiles of \oi\, 63 \um, and find self-absorption profiles for a total extinction $A_V \geq 5$ mag.  As the extinction increases, the self-absorption becomes more pronounced.  This result indicates that, in addition to the foreground ``skin'' layer, the well-shielded interior molecular region of the cloud can also contribute to the \oi\, optical depth and the subsequent self-absorption profile.  Indeed, in our models without an embedded PDR, we also find that the \oi\, optical depth increases from 1.8, to 4.0, to 4.9 as the extinction grows from $A_V = 1.5$, to 5.0, to 20.0 mag.  This increase must be due to the addition of interior, well-shielded molecular gas, where the O$^0$ abundance is low, but still large enough to contribute to the integrated optical depth.  

\cite{Goldsmith2023} extend this work by modeling W3 as consisting of two zones: a dense component ($n = 5 \times 10^5$   \cmc) associated with the \hii\, region, and a lower density component ($n = 250$   \cmc) associated with the foreground absorbing gas.  Because the foreground gas is offset in velocity from the dense component, the resulting self-absorbed line shapes are asymmetric.  The total extinction of the extended, foreground component is $A_V = 10$ mag.  They conclude that lower density foreground clouds are responsible for the observed \oi\, self-absorption features often seen toward high mass star forming regions.

Our results generally agree with those of \cite{Goldsmith2021, Goldsmith2023}. One important difference is that the foreground absorbing cloud toward W3 has a much larger column density than the foreground cloud associated with the Nessie Bubble.  Our results, however, also show that \oi\, will still exhibit self-absorption or absorption features even in the absence of a significant interior molecular zone, and absorption features will arise even in the absence of an embedded PDR.  Moreover, the ``skin'' hypothesis can also explain \cii\, absorption and self-absortion.  Given the very low abundances of \cii\, in the molecular interior of an opaque cloud, the molecular interior of a cloud cannot contribute to \cii\, absorption of self-absorption, as the \cii\, optical depths there are negligible.  The externally illuminated PDR ``skin,'' however, can plausibly produce the observed \cii\, absorption and self-absorption, at least in some cases.

\subsection{\rm{\oi\,} absorption associated with the Nessie Nebula}

In addition to the \cii\, and \oi\, absorption features associated with foreground clouds, the \oi\, spectrum toward AGAL337.916-00.477 also shows absorption features at velocities associated with molecular gas in the Nessie Nebula, specifically a broad feature at $v = -35$ \kms\, and another weaker feature at $v = -42$ \kms\, (see Fig. \ref{fig:figcontprofiles}).  These features are likely due to photodissociated gas on the exteriors of molecular clouds.  At the same position \hcop\, reveals an inverse P Cygni profile, with blueshifted emission and redshifted absorption (see Fig. \ref{fig:figposprofiles}, position 7). Such inverse P Cygni profiles are usually interpreted to indicate collapse, and may indicate the graviational collapse of the YSO core.   Because the redshifted HCO$^+$ absorption feature matches the velocity of the broad \oi\, absorption feature, the \oi\, absorption may well arise in the infalling exterior of the graviationally collapsing YSO core.  The weaker absorption feature has a velocity that matches the systemic velocity of the molecular gas associated with the Nessie Bubble, as evidenced by CO emission.  Thus, the weaker \oi\, absorption feature may arise in the molecular gas shell surrounding the Nessie Bubble.

\section{\cii\, high-velocity gas distribution toward the Nessie Bubble} 

The average \cii\ spectrum of the nebula in Fig. \ref{fig:figavgprofiles} shows very prominent high-velocity wings that reach velocities of 20 km s$^{-1}$, and potentially even up to 30 km s$^{-1}$, relative to the bulk velocity of the filament. These \cii\, high-velocity wings thus reach several times the escape velocity of the host molecular cloud. Similar high-velocity \cii\, features have been identified in nearly every \hii\, region that was observed with upGREAT \citep[e.g.,][]{Simon2012,Schneider2018,Pabst2019,Luisi2021,Tiwari2021,Beuther2022,Bonne2022a,Tram2023}. They have been associated with expanding bubbles in Orion and RCW 120 \citep[e.g.,][]{Pabst2019,Luisi2021}, but in other regions the high-velocity gas has a more complex morphology. In these complex regions, the \cii\, high-velocity gas is too concentrated at the projected PDR interface of the \hii\, region and molecular cloud to be consistent with a spherical expanding bubble or shell \citep{Kabanovic2022,Bonne2023a}. In some of these regions the high-velocity gas even extends beyond the PDR interface into the surrounding ISM \citep[e.g.,][]{Tiwari2021,Bonne2022a}. 

The same appears to be true for the Nessie Nebula. Fig. \ref{fig:ciiredblue} indicates the spatial distribution of the high-velocity ionized carbon gas in the region. The redshifted high-velocity ionized carbon gas follows the bright PDR ``question mark'' morphology quite closely, but then in the south of the map it extends significantly beyond the identified PDR interface (shown with the grayscale background image in Fig. \ref{fig:ciiredblue}). On the other hand, the blueshifted high-velocity ionized carbon gas follows the outline of the \hii\, regions fairly closely in the south, but in the north of the region, the blueshifted gas extends beyond the ``question mark'' PDR into an additional, but fainter, bubble. Indeed, the blueshifted \cii\, emission extends even beyond this additional bubble into the surrounding ISM.

Both blue- and redshifted ionized carbon gas are thus concentrated at the projected PDR interface and even extend beyond the bubble morphology, where they have a strikingly different spatial distribution. This morphology has a straightforward explanation if the bubble has broken open at several locations, which has been proposed for a few other regions as well \citep[e.g.,][]{Tiwari2021,Bonne2022a}.  Such gaps in the bubble allow the high-velocity gas to stream along the \hii\, region's walls out of the cloud and into the diffuse ISM, leading to a progressive erosion of the filament and ambient molecular cloud. The strikingly different spatial distribution of the blue- and redshifted gas is then likely associated with projection effects of the imhomogeneous cloud and filament morphology which leads to the bubble breaking open in specific directions. For example, in the southern region, the molecular cloud is likely mostly in front of the \hii\, region when viewed from earth, and thus exhibits blueshifted gas at the projected PDR interface of the \hii\, region and molecular cloud, while the redshifted gas can more easily flow beyond the projected PDR interface.

\section{Triggered Star Formation in the Dense IRDC Filament}

\begin{figure}
\begin{center}
\includegraphics[scale=0.5, angle=0]{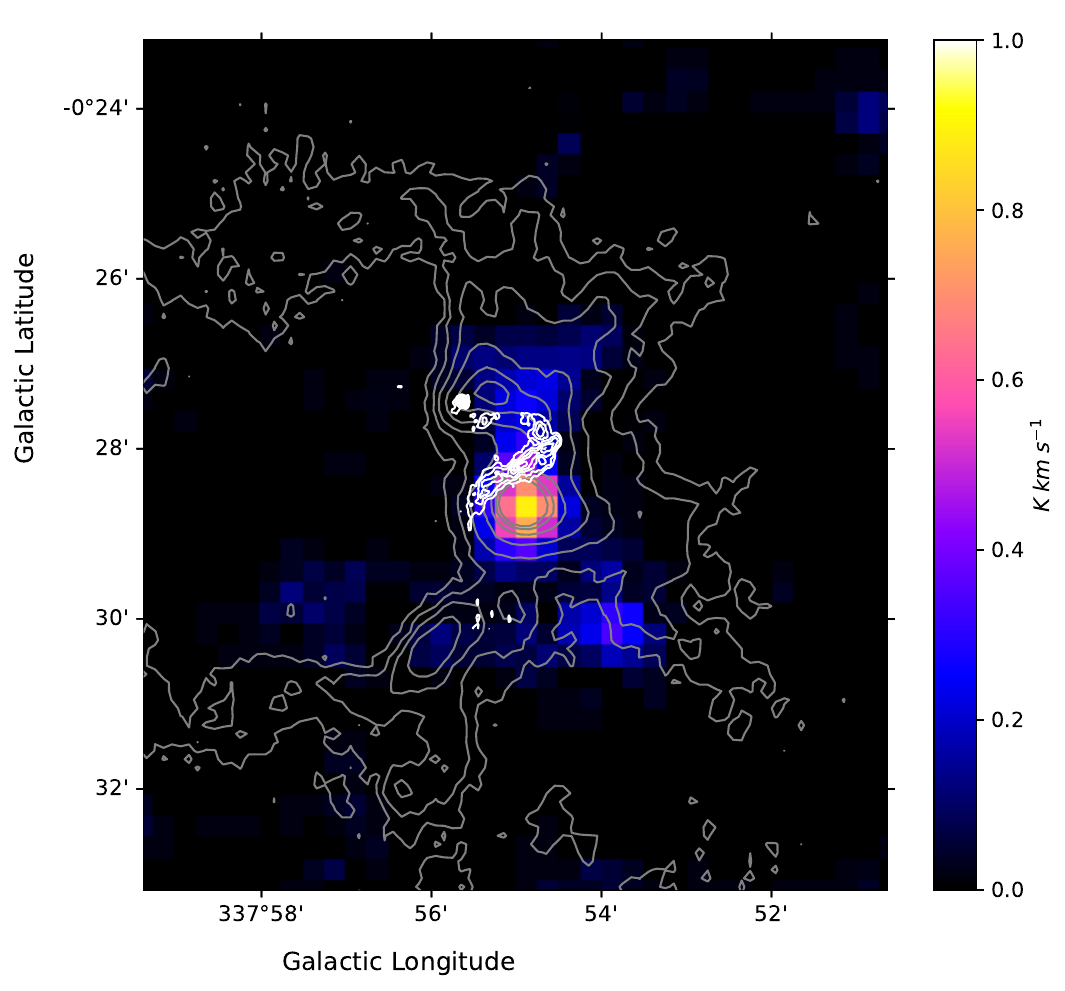} 
\includegraphics[scale=0.38, angle=0]{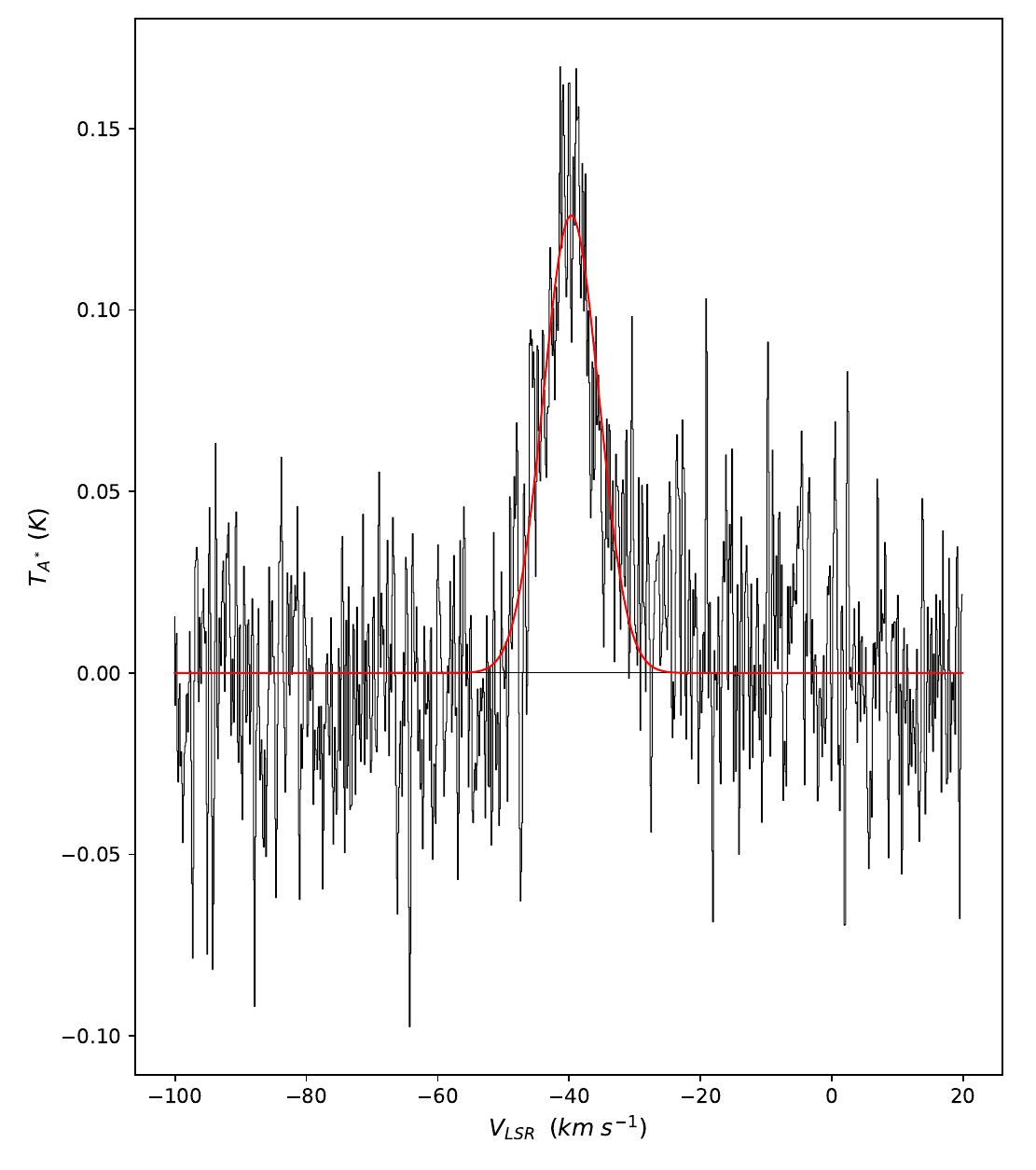} 
\includegraphics[scale=0.5, angle=0]{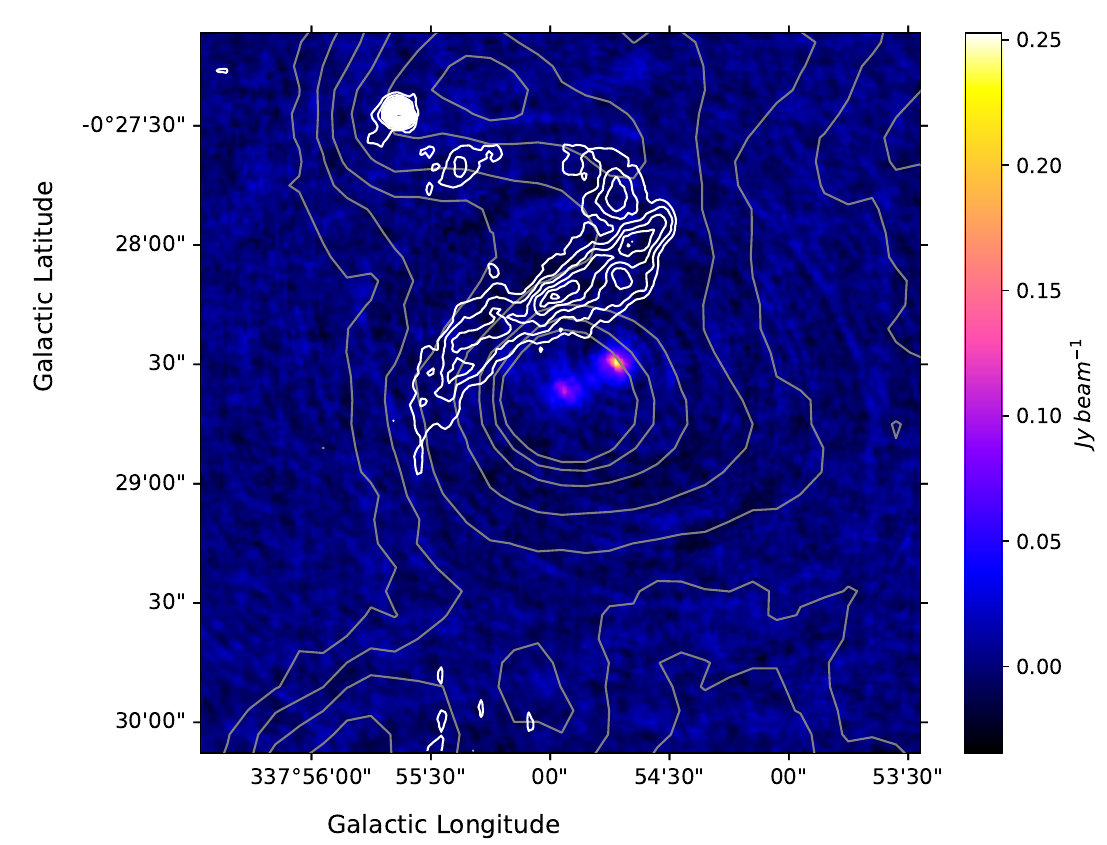}
\includegraphics[scale=0.38, angle=0]{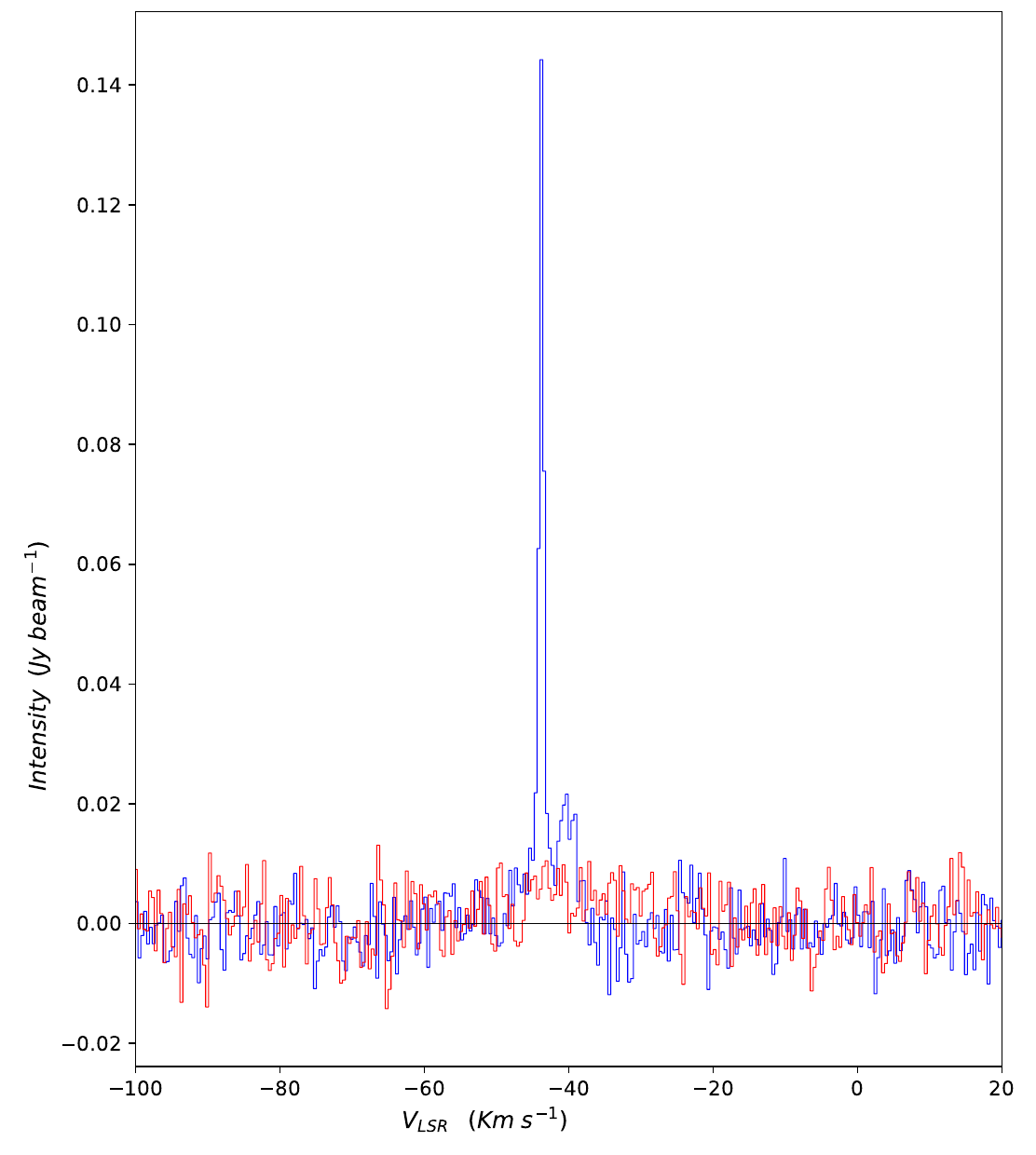}
\caption{(Top Left) SiO $2-1$ integrated intensity from Mopra (color), 24 GHz continuum from ATCA (white contours), and 870 \um\, continuum from ATLASGAL (gray contours) toward the Nessie Bubble. The contour levels are as follows.  White contours: 24 GHz continuum from this work, from 0.002 to 0.05 Jy beam$^{-1}$ in steps of 0.002Jy beam$^{-1}$; gray contours:  ATLASGAL 870 \um\, continuum, at 0.2, 0.5, 1.0, 2.0, 4.0, 6.0, and 8.0 Jy beam $^{-1}$. (Top Right) The SiO $2-1$ spectrum at the location of AGAL337.916-00.477.  The red line shows a Gaussian fit with a velocity dispersion $\sigma_V$ = 4.23 \kms. (Bottom Left) \nht\, (3,3) image of a single velocity channel containing the peak maser emission (\vlsr\, $= -38$ \kms).  White and gray contours are the same as in the top right panel.  (Bottom Right) The \nht\, (1,1) (red) and (3,3) (blue) spectrum toward the peak in the channel map.  The maser nature of the (3,3) emission is evident:  it is a point source; the line is spectrally unresolved ($\Delta V < 0.4$ \kms); and the (3,3) line is much brighter than any (1,1) emission at the same location.}
\label{fig:NessieA-SiO}
\end{center}
\end{figure}

The luminous YSO AGAL337.916-00.477, the site of most recent high-mass star formation in the region, lies exactly where the Nessie IRDC filament intersects the Nessie Bubble.  This location suggests that the interaction between the expanding Nessie Bubble with the dense filament may have triggered star fomation there.  Evidence for such an interaction comes from the presence of shock tracers toward the core.  For example, the SiO 2-1 image taken with the Mopra telescope shows a strong peak at the location of the YSO (see Figure \ref{fig:NessieA-SiO}).  Moreover, a comparison of the \nht\, (1,1) and (3,3) data near the YSO show a compact source of (3,3) emission in the absence of (1,1) emission.  The narrow (3,3)  linewidth at this location and the lack of any other associated \nht\, emission from any other inversion transitions indicate emission from a  \nht\, (3,3) maser (see Figure \ref{fig:NessieA-SiO}).  Although rare,  \nht\, (3,3) masers are associated with shocks in star forming regions (e.g., \citealt{Kraemer1995}) or in one case, an interaction between an expanding supernova shell and an IRDC filament \citep{Hogge2019}.   The observations thus suggest a strong interaction between the expanding Nessie Bubble and the Nessie IRDC filament at the position of the luminous YSO. Since the Nessie IRDC already contains dense gas and a large linear mass density, it is entirely plausible that further compression of the filament by the expanding Bubble may well have induced gravitational collapse that led to the formation of the AGAL337.916-00.477 YSO.

\begin{figure}
\begin{center}
\includegraphics[scale=1.0, angle=0]{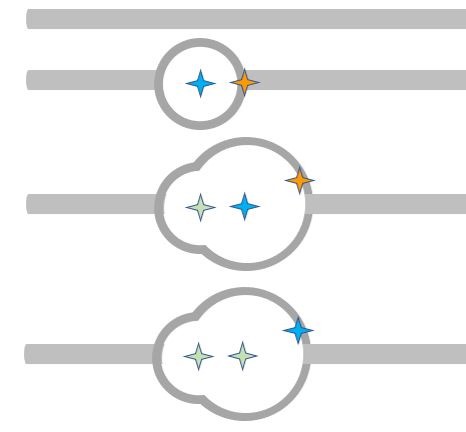}
\caption{A speculative scenario on propagating star formation along the Nessie Infrared Dark Cloud filament.  Here the gray indicates molecular gas, the blue stars a current location of high-mass, main sequence stars or star clusters, the green stars an evolved, inactive location of previous star formation, and the yellow starts the location of current pre-stellar high-mass star formation.  The interaction between the filament and the expanding bubbles triggers star formation. }
\label{fig:BubbleEvolution}
\end{center}
\end{figure}

\begin{figure}
\begin{center}
\includegraphics[scale=1.0, angle=0]{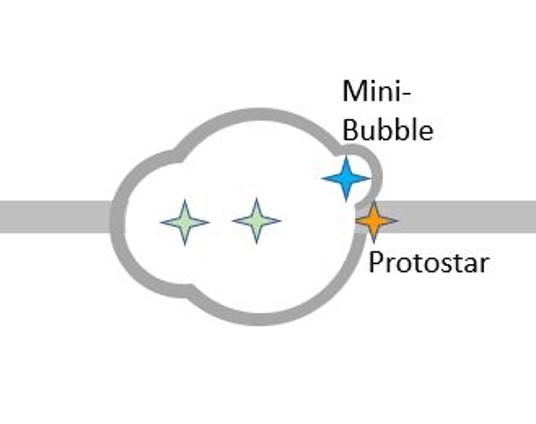}
\caption{A cartoon sketch of the current morphology of the Nessie Bubble.}
\label{fig:BubbleEvolution2}
\end{center}
\end{figure}

We hypothesize that the interaction between expanding \hii\, region bubbles and the Nessie IRDC filament, and the subsequent triggering of star formation, might explain some of the observed characteristics of the Nessie Bubble.  Figure \ref{fig:BubbleEvolution} shows a speculative scenario.  It begins with a single OB star or star cluster that forms in a filament and subsequently drives an \hii\, region bubble.  At a later time, the bubble interacts with the filament and forms a new OB cluster.  The new \hii\, region bubble expands into the already evacuated region formed from the first bubble.  The process repeats and a new cluster forms.  As the bubbles merge they collect a large amount of material on their periphery.  This material can then form new stars via the ``collect and collapse'' scenario \citep{ElmegreenLada1977, Zavagno2010}.  The final configuration (Figure \ref{fig:BubbleEvolution2}) resembles the broad outlines of the current morphology of the Nessie Bubble.  The luminous protostar AGAL337.916-00.477 is the latest protostar to form as the bubble triggers star formation in the filament.   The ``mini-bubble'' at the crook of the ``question mark'' is ionized and inflated by a previously formed OB star cluster, which also provides the ultraviolet radiation inside the merged bubble.  Finally, an ultracompact \hii\, region, revealed by the compact  radio continuum source to the northeast (at the end of the ``question mark''), is a slightly older YSO that has reached the main sequence and begun to ionize its surrounding medium.

The above scenario suggests that star formation can propagate along IRDC filmaments by triggering star formation via a series of expanding \hii\, region bubbles.  It is similar to previous scenarios of propagating star formation, except that the dense filament provides a natural pathway for star formation to proceed in successive cluster-forming events.  Of course, other explanations, including a large density gradient that could lead to an asymmetric shell expansion, could also account for the morphology of the Nessie Bubble.  The strong evidence for triggered star formation by the bubble-filament interaction, however, does suggest that this mechanism could provide a feedback mechanism that could be sustained in Nessie or other large IRDC filaments.

\section{Summary}
We present SOFIA FIR, ATCA radio, and Mopra mm observations of an \hii\, region bubble associated with the Nessie Nebula, a filamentary Infrared Dark Cloud.  SOFIA imaged the \cii\, and \oi\, lines, important probes of photodissociation regions.  ATCA imaged \nht\, (1,1) and (3,3) inversion lines to probe the molecular gas, as well as free-free radio continuum to probe the ionoized gas.  The Mopra telescope imaged the SiO $2-1$ line, a probe of shocked molecular gas.
From these data, we draw the following conclusions:

1. The overall structure of the western edge of the Bubble is consistent with expectations for an internally illuminated photodissociated region, ionized by the star cluster located just to the interior of the bubble.  The gas distribution takes the general shape of a question mark, with the cluster located inside the crook.  Ionized gas traced by radio continuum lies to the interior of the Bubble, followed by \cii\, and \oi\, emission tracing the photodissociated gas, and \nht\, emission toward the exterior of the Bubble.  The Nessie IRDC filament protrudes from the edge of the bubble to the west.

2. Toward the bright PDR line emission, the \oi\, line profiles show asymmetries and flat-topped shapes that indicate self-absorption throughout the region.  The most likely location for the self-absorbing gas is in the Bubble wall or the IRDC filament.

3. Toward the bright FIR/submm continuum source AGAL337.916-00.477, both \cii\, and \oi\, show an absorption feature at $-18$ \kms, corresponding to H I absorption and faint CO emission.  This absorption arises in an unrelated foreground molecular cloud with no sign of star formation or internal PDRs.  Absorption in \cii\, is marginally detected in another foreground cloud at $+3$ \kms.

4. Since the \cii\, and \oi\, absorbing gas at \vlsr $= -18$ \kms\, is not associated with star formation,  we suggest that it is associated with the periphery of a foreground cloud exposed to a standard interstellar radiation field with $G_0 \sim 1$.  The periphery of such a cloud satisfies both conditions for producing an absorption feature: (1) a low excitation temperature ($T_{ex} \lesssim 20$ K), and (2) sufficient optical depth ($\tau \gtrsim 1$).  The low excitation temperature is due to the fact that the gas densities in typical giant molecular clouds are far below the critical densities for the \cii\, and \oi\, lines, and thus the excitation is subthermal.  External illumination by a standard intersellar radiation field can produce a photodissociated exterior with sufficient column densities of C$^+$ and O$^0$ to produce enough optical depth to absorb the background continuum signal.  Both rough estimates and detailed modeling using the Meudon PDR code confirm that both conditions are satisfied. Specifically, for the $v = -18$ \kms\, absorption feature toward AGAL337.916-00.477, the Meudon model predictions for a foreground absorbing cloud with $A_V = 1.6$ mag matches the observed CO, \cii, and \oi\, data well.  In general, the exteriors of externally illuminated, foreground molecular clouds can account for the ubiquity of \cii\, and \oi\, absorption and self-absorption features.  Such features are significant because the derived parameters from PDR models depend sensitively on the \cii\, and \oi\, fluxes.  If absorption or self-absorption significantly diminishes their fluxes, the uncorrected fluxes will lead to erroneous estimates of densities and UV field strengths.

5.  The luminous YSO AGAL337.916-00.477 is located precisely at the location where the Bubble and the IRDC filament intersect.  The presence of two shock tracers, SiO $2-1$ and \nht\, (3,3) maser emission, at this location strongly indicates an interaction between the expanding Bubble and the IRDC filament.  The YSO's formation is likely to have been triggered by this interaction.

6. If Bubble-Filament interactions have triggered star formation in the past, star formation may well have propagated along the filament.  Past episodes of star formation might explain the current morphology of the Bubble, including the location of the star cluster, the ``mini-bubble'' extension of the main bubble near the cluster, and the location of the luminous YSO.

\section{Acknowledgements} 
The Stratospheric Observatory for Infrared Astronomy (SOFIA) was jointly operated by the Universities Space Research Association, Inc. (USRA), under NASA contract NAS2-97001, and the Deutsches SOFIA Institut (DSI) under DLR contract 50 OK 0901 to the University of Stuttgart.     We thank the upGREAT team for their support in observing and calibrating the data. The development and operation of GREAT were financed by resources from the MPI für Radioastronomie (Bonn), Universit\"at zu K\"oln, the DLR Institut für Optische Sensorsysteme, Berlin, the Deutsche Forschungsgemeinschaft (DFG, German Research Foundation) within the Collaborative Research Center 956, and the Federal Ministry of Economics and Energy (BMWI) via the German Space Agency (DLR) under grants 50 OK 1102, 50 OK 1103, and 50 OK 1104. R. Simon gratefully acknowledges support within the Collaborative Research Center 1601 (SFB 1601 sub-project B2) funded by the Deutsche Forschungsgemeinschaft (DFG) – project ID 500700252.  The Australia Telescope Compact Array is part of the Australia Telescope National Facility (https://ror.org/05qajvd42) which is funded by the Australian Government for operation as a National Facility managed by CSIRO.  We acknowledge the Gomeroi people as the traditional owners of the ATCA Observatory site.  We thank Nigel Maxted for his assistance and support for the Mopra observations.  We thank Peter Barnes for assistance in obtaining ThrUMMS data.  We thank Xander Tielens for pointing out that the surface chemistry of water ice on dust mantles plays an important role in the atomic oxygen distribution in irradiated molecular clouds.  We thank Paul Goldsmith for hepful discussions on \oi\, excitation.  PDR models published in this paper have been produced with the Meudon PDR code, available at http://ism.obspm.fr.  We thank Franck Le Petit and Jacques Le Bourlot for their great assistance with implementing the Meudon PDR code, and for providing new surface chemistry parameters ahead of their public release.  Their help was invaluable in this work.  R. Simon acknowledges support from the Collaborative Research Centre 956, funded by the Deutsche Forschungsgemeinschaft (DFG) – project ID 184018867. P. Sanhueza was partially supported by a Grant-in-Aid for Scientific Research (KAKENHI Number JP22H01271 and JP23H01221) of JSPS. We thank an anonymous referee for making several important suggestions that have greatly improved this paper.

\facilities{SOFIA, ATCA, Mopra}

\software{CLASS/GILDAS (http://www.iram.fr/IRAMFR/GILDAS), Astropy \citep{Astropy2022},  Miriad \citep{Sault1995}}

\newpage

\end{document}